\documentclass[11pt,a4paper]{article}
\pdfoutput=1
\usepackage{jheppub}
\usepackage{amsfonts}
\usepackage{amsmath}
\usepackage{cancel}
\usepackage{graphicx}
\usepackage{hyperref}
\usepackage[utf8]{inputenc}
\usepackage{multirow}
\usepackage[table]{xcolor}
\usepackage{xspace}
\usepackage{physics}
\usepackage{diagbox}
\usepackage[T1]{fontenc}
\usepackage[font=small,labelfont=bf,tableposition=top]{caption}

\definecolor{MyGreen}{HTML}{00a85c}

\newcommand{\tp}[1]{{\color{red}  #1}}

\newcommand{\HEPfit}{\texttt{HEPfit}\xspace}

\hypersetup{pdfstartview=FitV,colorlinks=true,linkcolor=blue,citecolor=red,filecolor=black,urlcolor=blue}

\definecolor{sigstrcolor1}{HTML}{ffffff}
\definecolor{sigstrcolor2}{HTML}{ffffff}
\definecolor{sigstrcolor3}{HTML}{ffffff}
\definecolor{sigstrcolor4}{HTML}{ffffff}
\definecolor{sigstrcolor5}{HTML}{ffffff}
\definecolor{sigstrcolor6}{HTML}{ffffff}
\definecolor{color1}{HTML}{00dd00}
\definecolor{color2}{HTML}{dd0000}
\definecolor{color3}{HTML}{000000}
\definecolor{color4}{HTML}{000000}
\definecolor{color5}{HTML}{000000}
\definecolor{color6}{HTML}{000000}
\definecolor{color7}{HTML}{000000}
\definecolor{color8}{HTML}{000000}
\definecolor{color9}{HTML}{000000}
\definecolor{color10}{HTML}{000000}
\definecolor{color11}{HTML}{000000}
\definecolor{color12}{HTML}{000000}

\definecolor{color8TeV}{HTML}{ddddff}
\definecolor{color13TeV}{HTML}{eeee99}
\definecolor{color2TeV}{HTML}{9DCEBF}
\DeclareCaptionLabelFormat{andtable}{#1~#2  \&  \tablename~\thetable}

\def\beqn{\begin{eqnarray}}
\def\eeqn{\end{eqnarray}}
\def\cZ{{\cal Z}}

\newcommand{\comOUT}[1]{{}}

\begin{document}

\title{Global fits in the Aligned Two-Higgs-Doublet model}

\author[a]{Otto Eberhardt,}
\emailAdd{otto.eberhardt.physics@gmail.com}

\author[b]{Ana Pe\~nuelas Mart\'inez,}
\emailAdd{apenuela@uni-mainz.de}

\author[a]{Antonio Pich}
\emailAdd{antonio.pich@ific.uv.es}

\affiliation[a]{Instituto de F\'isica Corpuscular, Parque Cient\'ifico, 
C/Catedr\'atico Jos\'e Beltr\'an, 2, E-46980 Paterna, Spain}

\affiliation[b]{PRISMA+ Cluster of Excellence \& Mainz Institute for Theoretical Physics, Johannes Gutenberg University, 55099 Mainz, Germany}

\date{\today}

\abstract{
We present the results of a global fit to the Aligned Two-Higgs Doublet Model, assuming that there are no new sources of CP violation beyond the quark mixing matrix. We use the most constraining flavour observables, electroweak precision measurements and the available data on Higgs signal strengths and collider searches for heavy scalars, together with the theoretical requirements of perturbativity and positivity of the scalar potential. The combination of all these constraints restricts the values of the scalar masses, the couplings of the scalar potential and the flavour-alignment parameters.
The numerical fits have been performed using the open-source \HEPfit package.
}

\preprint{{\raggedleft IFIC/20-53 \\ MITP-20-082 \par}}

\maketitle

\section{Introduction}
\label{sec:intro}

The particle content of the Standard Model (SM) was confirmed in 2012 with the discovery of the Higgs boson \cite{Aad:2012tfa, Chatrchyan:2012xdj}. However there are still some aspects of Nature that can not be explained by the SM alone and leave room for new physics (NP). In particular, scalars transforming as doublets under the $SU(2)_L$ group, and therefore satisfying the successful mass relation $M_W = M_Z \cos \theta$, are appropriate candidates for building extended electroweak models.

The simplest of these extensions is the two-Higgs doublet model (2HDM), containing a second Higgs doublet with the same quantum numbers as the SM one. 
In order to avoid dangerous flavour-changing neutral-current (FCNC) transitions, the usual implementations of the model \cite{Gunion:1989we, Branco:2011iw,Ivanov:2017dad} assume specific 
discrete $\mathcal{Z}_2$ symmetries to constrain the Yukawa sector~\cite{Glashow:1976nt}, so that only one scalar doublet can couple to a given right-handed fermion. However, to comply with the strong phenomenological constraints, it is enough to impose the much weaker assumption of flavour alignment \cite{Manohar:2006ga,Pich:2009sp,Pich:2010ic,Penuelas:2017ikk}, \emph{i.e.}, that the Yukawa couplings of the two scalar doublets have the same flavour structure. This leads to a more general framework with minimal flavour violation \cite{Chivukula:1987py,DAmbrosio:2002vsn}, where FCNC  couplings are absent at tree-level and very suppressed at higher orders \cite{Pich:2009sp,Jung:2010ik,Braeuninger:2010td,Bijnens:2011gd,Penuelas:2017ikk}.

Constraints on the 2HDM parameters have been widely studied, taking into account recent LHC data \cite{Chen:2013kt, Celis:2013rcs,Chiang:2013ixa, Grinstein:2013npa, Barroso:2013zxa, Xu:2013ela, Eberhardt:2013uba, Celis:2013ixa, Chang:2013ona, Wang:2013sha, Baglio:2014nea, Inoue:2014nva, Kanemura:2014bqa, Ferreira:2014qda, Dumont:2014lca, Bernon:2014nxa, Chowdhury:2015yja, Craig:2015jba, Bernon:2015qea, Bernon:2015wef, Cacchio:2016qyh, Belusca-Maito:2016dqe, Sanyal:2019xcp, Herrero-Garcia:2019mcy,  Karmakar:2019vnq,Chen:2019pkq,Arco:2020ucn,Rajec:2020orn,Chen:2020aht, Aiko:2020ksl}, together with other requirements from flavour and LEP physics, and theoretical considerations. However these analyses normally considered specific 2HDM models with $\mathcal{Z}_2$ symmetries \cite{ Eberhardt:2018lub, Chowdhury:2017aav, Eberhardt:2017ulj, Cacchio:2016qyh, Eberhardt:2015ypa, Chowdhury:2015yja,Eberhardt:2014kaa,Eberhardt:2013uba, Gunion:2002zf}. 
In this work we have performed a global fit to the relevant experimental and theoretical constraints in the much more general framework of the (flavour) aligned two-Higgs doublet model (A2HDM) \cite{Pich:2009sp,Pich:2010ic}. To simplify the analysis, we have neglected additional sources of CP violation beyond the quark mixing matrix, {\it i.e.}, we assume a CP-conserving scalar potential and real alignment parameters. 

This paper is organized as follows: Section~\ref{sec:model} contains a brief overview of the model. In Section~\ref{sec:2HDMFitConstraints} the fit  setup and the theoretical and experimental constraints considered are explained. The results of the fit are presented in Sections \ref{sec:FitResults_Light} and~\ref{sec:FitResults_Heavy}, which discuss the two possible mass orderings for the observed 125~GeV Higgs, being either the lightest CP-even scalar or the heaviest one. Our main conclusions are finally given in Section~\ref{sec:summary}. An appendix compiles the collider data sources employed in our global fit.

\section{The Aligned Two-Higgs-Doublet model}
\label{sec:model}

Let us consider the SM extended with a second complex scalar doublet of hypercharge $Y=~\frac{1}{2}$. In general, the neutral components of both doublets can acquire vacuum expectation values. However, making a global $SU(2)$ transformation in the scalar space spanned by the two doublets, it is always possible to work in the so-called Higgs basis,
\begin{equation}
\Phi_1 =   \begin{bmatrix} G^+ \\
\frac{1}{\sqrt{2}}\, (v + S_1 + i\, G^0) \end{bmatrix},  
\qquad\qquad 
\Phi_2 =   \begin{bmatrix} H^+ \\
\frac{1}{\sqrt{2}}(S_2 + i\, S_3) \end{bmatrix},  
\label{eq:doublets}
\end{equation}
where only one doublet has non-zero vacuum expectation value, with $v= (\sqrt{2}\, G_F)^{-1/2}\approx 246$~GeV. The field $\Phi_1$ plays the role of the SM Higgs doublet with $G^0$ and $G^{\pm}$ the electroweak Goldstone bosons. The scalar spectrum contains five degrees of freedom: the charged scalars $H^{\pm}$ and three neutral fields $S_i$.

The most general scalar potential, invariant under $SU(2)_L\otimes U(1)_Y$ reads
\begin{align}
 V
 & = \mu_1\,\Phi_1^\dagger\Phi_1^{\phantom{\dagger}}
   + \mu_2\,\Phi_2^\dagger\Phi_2^{\phantom{\dagger}}
   + \left[\mu_3\, \Phi_1^\dagger\Phi_2^{\phantom{\dagger}}
              +\mu_3^*\,\Phi_2^\dagger\Phi_1^{\phantom{\dagger}}\right]
\nonumber \\ &              
   + \tfrac12 \lambda_1\left(\Phi_1^\dagger\Phi_1^{\phantom{\dagger}}\right)^2
   + \tfrac12 \lambda_2\left(\Phi_2^\dagger\Phi_2^{\phantom{\dagger}}\right)^2 
  + \lambda_3 \left(\Phi_1^\dagger\Phi_1^{\phantom{\dagger}}\right)
            \left(\Phi_2^\dagger\Phi_2^{\phantom{\dagger}}\right)
  + \lambda_4 \left(\Phi_1^\dagger\Phi_2^{\phantom{\dagger}}\right)
            \left(\Phi_2^\dagger\Phi_1^{\phantom{\dagger}}\right)
\nonumber \\  &        
  + \left[ \left(\tfrac12  \lambda_5\, \Phi_1^\dagger\Phi_2^{\phantom{\dagger}}
  + \lambda_6 \, \Phi_1^\dagger\Phi_1^{\phantom{\dagger}} 
   +\lambda_7\,  \Phi_2^\dagger \Phi_2^{\phantom{\dagger}} \right) \left( \Phi_1^\dagger\Phi_2^{\phantom{\dagger}} \right)
                      + {\rm h.c.} \right] ,
\label{eq:V2hdm}
\end{align}
where all parameters are real except $\mu_3$, $\lambda_5$, $\lambda_6$ and $\lambda_7$. The minimization of the potential (in the Higgs basis) gives the relations  $\mu_1 = -\frac{1}{2}\,\lambda_1 v^2$ and $\mu_3 = -\frac{1}{2}\,\lambda_6 v^2$. Moreover, one phase can be reabsorbed into the field $\Phi_2$. Thus, the potential is fully characterized by eleven real parameters: $v$, $\mu_2$, $\lambda_{1,2,3,4}$, $|\lambda_{5,6,7}|$,
and the two relative phases between $\lambda_5$, $\lambda_6$ and $\lambda_7$.
To simplify the analysis, we will assume a CP-conserving potential with all couplings real, which reduces the number of degrees of freedom to nine.

The quadratic terms in the potential determine the physical scalar masses \cite{Celis:2013rcs}:
\begin{equation}\label{eq:MAHpm}
M^2_{H^\pm} = \mu_2 +\frac{1}{2}\,\lambda_3\, v^2\, ,
\qquad\qquad\quad
M^2_{A} = M^2_{H^\pm} +\frac{1}{2} \left(\lambda_4-\lambda_5\right) v^2\, ,
\end{equation}
\begin{equation}\label{eq:MhH}
M^2_{h} = \frac{1}{2} \left( \Sigma -\Delta\right) ,
\qquad\qquad\qquad
M^2_{H} = \frac{1}{2} \left( \Sigma +\Delta\right) ,
\end{equation}
where
\begin{equation}
\Sigma \, =\, M^2_{H^\pm} + \frac{1}{2}\, v^2 \left( 2 \lambda_1 + \lambda_4 + \lambda_5\right) ,
\end{equation}
and
\begin{equation}
\Delta \, =\, \sqrt{\left[ M^2_{H^\pm} + \frac{1}{2}\, v^2 \left(-2 \lambda_1 + \lambda_4 + \lambda_5\right)\right]^2 + 4 v^4 (\lambda_6)^2}\, .
\end{equation}
$A = S_3$ is a CP-odd neutral scalar, while the CP-even neutral mass eigenstates are linear combinations of $S_1$ and $S_2$,
\begin{equation}\label{eq:mixtureCP}
 \left( \begin{array}{c} h \\ H  \end{array}\right)\, =\,
\left[ 
\begin{array}{cc}  \cos{\tilde{\alpha}} & \sin{\tilde{\alpha}} \\
- \sin{\tilde{\alpha}} & \cos{\tilde{\alpha}} \\  \end{array}
\right]
\left( \begin{array}{c} S_1 \\ S_2  \end{array}\right) ,
\end{equation}
with
\begin{equation}
\tan{\tilde{\alpha}}\, =\, \frac{M_h^2-\lambda_1 v^2}{v^2\lambda_6}\, =\, \frac{v^2\lambda_6}{\lambda_1 v^2- M_H^2}\, .
\end{equation}

The couplings of a single neutral scalar with a pair of gauge bosons are identical to the SM ones, with the field $S_1$ taking the role of the SM Higgs. Therefore ($VV=W^+W^-,ZZ$),
\begin{equation}\label{eq:gaugecouplings}
g_{hVV} \, =\, \cos{\tilde{\alpha}}\;  g_{hVV}^{\mathrm{SM}}\, ,
\qquad\qquad
g_{HVV} \, =\, -\sin{\tilde{\alpha}}\;  g_{hVV}^{\mathrm{SM}}\, ,
\qquad\qquad
g_{AVV} \, =\, 0\, .
\end{equation}
The complete list of gauge couplings and scalar interactions can be found in Ref.~\cite{Celis:2013rcs}. 

In terms of fermion mass eigenstates the Yukawa Lagrangian reads:
\begin{eqnarray}
\nonumber
\mathcal{L}_{\text{Yuk}} &=& - \left( 1+\frac{S_1}{v} \right)\left\lbrace \bar{d}_L M_d d_R + \bar{u}_L M_u  u_R +\bar{\ell}_L M_{\ell} \ell_R \right\rbrace \\
&& -\,\frac{1}{v} \left(S_2 +i S_3 \right) \left\lbrace \bar{d}_L Y_d d_R + \bar{u}_L Y_u  u_R +\bar{\ell}_L Y_{\ell} \ell_R \right\rbrace \\
&& -\, \frac{\sqrt{2}}{v} H^{+} \left\lbrace \bar{u}_L V_{\text{CKM}}Y_d d_R -  \bar{u}_R Y_u^{\dagger} V_{\text{CKM}} d_L +  \bar{\nu}_L Y_{\ell} \ell_R \right\rbrace + \text{h.c.} \, ,
\nonumber
\label{eq:YukL1}
\end{eqnarray}
where all fermionic fields are written as 3-dimensional flavour vectors, $M_f$ ($f=d,u,\ell$) are the diagonal mass matrices and
$V_{\mathrm{CKM}}$ is the usual Cabibbo-Kobayashi-Maskawa (CKM) quark-mixing matrix.
In general, the Yukawa matrices $Y_f$ of the second doublet are not related to the fermion mass matrices and their elements can take arbitrary values, yielding FCNCs which are tightly constrained phenomenologically \cite{Pich:2018njk}.  
The dangerous FCNC transitions can be easily avoided at tree level, imposing that
only a single flavour structure is present for each right-handed fermion, \emph{ i.e.}, that the Yukawa matrices are aligned in the flavour space \cite{Pich:2009sp},
\begin{equation}
Y_{d,\ell} = \varsigma_{d,\ell} \, M_{d,\ell}\, ,
\qquad\qquad
Y_{u} = \varsigma_{u}^* \, M_{u}\, ,
\label{eq:alignment}
\end{equation}
where $\varsigma_f$ are complex numbers called alignment parameters.
This alignment condition determines the Yukawa Lagrangian of the A2HDM  \cite{Pich:2009sp,Pich:2010ic,Manohar:2006ga}: 
\begin{eqnarray}
\mathcal{L}_{\text{Yuk}} &=& - \frac{\sqrt{2}}{v}\, H^+ \left\lbrace \bar{u} \left[ \varsigma_d V_{\mathrm{CKM}} M_d \mathcal{P}_R - \varsigma_u M_u^{\dagger} V_{\mathrm{CKM}} \mathcal{P}_L\right] d + \varsigma_\ell\, \bar{\nu} M_\ell \mathcal{P}_R \ell \right\rbrace  
\nonumber\\  
&& -\, \frac{1}{v}\, \sum_{i, f} 	y_{f}^{\varphi_i^0}\, \varphi_i^{0}\;  \left[\bar{f} M_f \mathcal{P}_R f \right] + \mathrm{h.c.} \, , 
\label{eq:Yukawa2}
\end{eqnarray}
where $\mathcal{P}_{R,L} = (1\pm\gamma_5)/2$ are the chirality projectors,
$\varphi_i^0 = h, H, A$ the scalar mass eigenstates and $y_f^{\varphi_i^0}$ their Yukawa couplings,
\begin{alignat}{5}
\nonumber
y_{d,\ell}^{h} &= \cos{\tilde{\alpha}} + \sin{\tilde{\alpha}}\; \varsigma_{d,\ell}\, ,  
\qquad\qquad
&y_{d,\ell}^{H} &= - \sin{\tilde{\alpha}} + \cos{\tilde{\alpha}} \; \varsigma_{d,\ell}\, ,  
\qquad\qquad
&y_{d,\ell}^{A} &= i\,\varsigma_{d, \ell}\, ,
\\ 
y_{u}^{h} &= \cos{\tilde{\alpha}} + \sin{\tilde{\alpha}}\; \varsigma_{u}^*\, ,  
\qquad\qquad
&y_{u}^{H} &= - \sin{\tilde{\alpha}} + \cos{\tilde{\alpha}} \; \varsigma_u^*\, ,  
\qquad\qquad
&y_{u}^{A} &= -i \, \varsigma_u^*\, . 
\label{eq:y_coup}
\end{alignat}
To simplify the analysis, we will assume real alignment parameters $\varsigma_f$. Thus, the only source of CP violation will be the CKM matrix.

The usual 2HDMs based on discrete $\mathcal{Z}_2$ symmetries are recovered by setting $\mu_3=\lambda_6=\lambda_7=0$, and correlating the alignment parameters through one of the following four possible choices: $\varsigma_{d} =\varsigma_{u} =\varsigma_{\ell} = \cot{\beta}$ (type I);
$\varsigma_{d} =\varsigma_{\ell} = -\tan{\beta},\ \varsigma_{u} =\cot{\beta}$ (type II);
$\varsigma_{d} =\varsigma_{u} = \cot{\beta},\ \varsigma_{\ell} = -\tan{\beta}$ (type X);
and
$\varsigma_{d} = -\tan{\beta},\ \varsigma_{u} =\varsigma_{\ell} =\cot{\beta}$ (type Y). The particular type-I model with $\cot{\beta}=0$ is known as inert 2HDM.

\section{Fit setup and constraints}
\label{sec:2HDMFitConstraints}

For our analysis we consider the CKM matrix as the only source of CP violation. Thus, we are assuming that the couplings of the scalar potential in Eq.~\eqref{eq:V2hdm} and the alignment parameters in Eq.~\eqref{eq:alignment} are real. 
The parameter space of the A2HDM is then characterized by twelve real quantities: the three alignment parameters and nine degrees of freedom in the scalar potential which we choose to be $v$, the four scalar masses, the CP-even mixing angle $\tilde\alpha$ and the quartic couplings $\lambda_{5,6,7}$. Two inputs are already empirically determined: the vacuum expectation value and the Higgs mass $m_h= 125.10\pm 0.14$~GeV \cite{Zyla:2020zbs}.\footnote{From now on we denote by $h$ the already discovered Higgs-like boson, and use $H$ for the second CP-even boson, irrespective of their mass ordering.}
The numerical values of the relevant SM parameters entering the fits are compiled in Table~\ref{tab:SM}.

\begin{table}[htb]
    \centering
    \begin{tabular}{|c|c| c || c | c | c | }
    \hline
    \textbf{Constant} & \textbf{Value} & \textbf{Ref.} & \textbf{Constant} & \textbf{Value} & \textbf{Ref.} \\
    \hline
    $G_F$ & $1.166\, 378\, 7\, (6) \cdot 10^{-5}\;\mathrm{GeV}^{-2}$ & \cite{Zyla:2020zbs} &  $m_t$ & 
    $172.4\, (7)\;\mathrm{GeV}$   
    & \cite{Zyla:2020zbs} \\
    $M_Z$ & $91.1876\, (21)\;\mathrm{GeV}$ & \cite{Zyla:2020zbs} &$m_b$ & $4.18\, (3)\;\mathrm{GeV}$ & \cite{Zyla:2020zbs} \\
    $\alpha$ &  $7.297\, 352\, 5693\, (11) \cdot 10 ^{-3}$    & \cite{Zyla:2020zbs} &  $m_c$ & $1.27\, (2)\;\mathrm{GeV}$ & \cite{Zyla:2020zbs} \\
    $m_h$ & $125.10\, (14)\;\mathrm{GeV}$   & \cite{Zyla:2020zbs} & $\alpha_s(M_Z)$ & $0.1179\, (10)$ & \cite{Zyla:2020zbs}\\
    $\Delta\alpha^{(5)}_{\mathrm{had}}(M_Z)$ & $0.02753 \pm 0.00010$ & \cite{Davier:2019can} &&&
    \\
    \hline
    \end{tabular}
    \caption{Numerical values for the SM parameters used in the fits. $m_b$ and $m_c$ denote the bottom and charm running quark masses, in the $\overline{\mathrm{MS}}$ scheme, at $\mu=2\;\mathrm{GeV}$, while $m_t$ is the value of the pole top mass extracted from cross-section measurements.}
        \label{tab:SM}
    \end{table}

Our fits have been performed with the open-source \HEPfit package \cite{deBlas:2019okz,hepfitsite},\footnote{The \HEPfit version used in this work corresponds to the \texttt{git} revision of 09/2020 with the choice of model class \texttt{GeneralTHDM}. The version used in this work is available at \cite{hepfitsite}. }
which uses a Markov-Chain Monte-Carlo implementation based on the Bayesian Analysis Toolkit~\cite{Caldwell:2008fw}. 
We assume the following priors for the fitted parameters:
$$
\abs{\lambda_{5,6,7}} < 10\, , 
\qquad \qquad 
\tilde{\alpha} \in \left[-\frac{\pi}{2}, \frac{\pi}{2}\right]\, ,
\qquad \qquad 
M_{A, H^{\pm}}^2 \in [10^2, 1500^2] \; \mathrm{GeV}^2\, ,
$$
\begin{equation}\label{eq:priors}
\varsigma_u \in [-1.5,1.5]\, , 
\qquad\qquad 
\varsigma_d \in [-50,50]\, , 
\qquad\qquad 
\varsigma_{\ell} \in [-100,100] \, .    
\end{equation}
The priors of the remaining CP-even scalar mass depend on the scenario studied. {\it Light (heavy) scenario} refers to the case in which the observed Higgs with a mass around 125 GeV ($h$) is the lightest (heaviest) CP-even scalar of the model. In this paper we will focus on
the {\it light scenario}, selected with the boolean flag \texttt{SMHiggs} set to \texttt{true},\footnote{Note that the \texttt{SMHiggs} flag is not listed in \cite{deBlas:2019okz}, since it was included after this documentation was released.} 
and adopt as mass priors for the non-SM Higgs ($H$)
\begin{equation}
 M_{H}^2 \in [125^2, 1500^2] \; \mathrm{GeV}^2\, .
\end{equation}
Nevertheless, in Section~\ref{sec:FitResults_Heavy} we will also discuss the implications of our fitted data set on
the {\it heavy scenario}, selected with the boolean flag \texttt{SMHiggs} set to \texttt{false},  
taking as mass priors
\begin{equation}
 M_{H}^2 \in [10^2, 125^2] \; \mathrm{GeV}^2\, .
\end{equation}
A more detailed analysis of the {\it heavy scenario}, including additional data from light scalar searches is deferred to a future work.
The scalar masses are chosen in a range such that they are relevant for the future LHC searches. The selected priors for the scalar potential parameters $\lambda_i$ are conservative, since larger values are excluded by theoretical constraints. The mixing angle $\tilde{\alpha}$ is varied in its full domain, and the alignment parameters $\varsigma_f \, (f = u,d,\ell)$ are varied within their perturbative ranges, \textit{i.e.}, $\sqrt{2}\, \varsigma_f  m_f/v \leq 1$. 

Bayesian statistics does not provide an unambiguous way to determine the prior distributions. A rule of thumb would be considering as flat priors the ones appearing linearly in our observables.  However, for the mass parameters this does not give a unique choice: while direct searches depend linearly on the heavy scalar masses, loop-induced processes appearing in flavour observables and in the Higgs signal strengths depend on the masses squared. To avoid a possible bias in the choice of these priors, we have performed fits with two different mass parametrizations. These two choices of mass priors are selected with the boolean flag  \texttt{use\_sq\_mass}. If it is set to \texttt{true} (\texttt{false}) squared (linear) mass priors are used. The effect of the choice of mass priors will be commented in the cases of interest. When the choice of the mass priors is irrelevant, squared mass priors will be used.

The global fit includes the theoretical and experimental constraints discussed below. Theoretical constraints, electroweak precision observables and  some of the Higgs constraints and direct searches were already included in the \texttt{THDM} model class and have been adapted to
the more general A2DHM case. The relevant flavour observables and the more recent experimental information on direct searches and Higgs signal strengths have been implemented in the \HEPfit package for this analysis.

\subsection{Theoretical constraints}
\label{sec:theoryconstraints}
To assure that the scalar potential is bounded from below, one must impose
the following positivity constraints on the quartic couplings $\lambda_i$ \cite{Deshpande:1977rw, Ferreira:2004yd}: 
\beqn
\notag
&&\lambda_1 \ge 0 \, , \qquad    \lambda_2 \ge 0 \, , 
\qquad    \sqrt{\lambda_1\lambda_2} + \lambda_3 \ge 0 \, ,
\qquad    \sqrt{\lambda_1\lambda_2} + \lambda_3 + \lambda_4 -|\lambda_5| \ge 0 \, , \\
&&\frac{1}{2}\left(\lambda_1 + \lambda_2\right) + \lambda_3 + \lambda_4 + \lambda_5 - 2\, |\lambda_6 + \lambda_7| \ge 0 \, .
\label{eq:theoLambda}
\eeqn
These necessary conditions restrict the allowed pattern of scalar masses. 

By imposing perturbative unitarity of the $S$-matrix we avoid that a given combination of parameters results in a too large scattering amplitude that violates the unitarity limit at a given perturbative order. Thus, we are actually requiring that the perturbative series does not break down. 
Here, unitarity is enforced for two-to-two scattering of scalar particles at leading order (LO), using \cite{Ginzburg:2005dt}

\begin{equation}
\left( a_j^{(0)} \right)^2 \leq \frac{1}{4}\, ,
\label{eq:unitarity}
\end{equation}
where $a_j^{(0)}$ are the tree-level contributions to the $j$ partial wave amplitude. For the high-energy scattering of scalars, only the $S$-wave amplitude ($j=0$) is relevant at LO. The corresponding matrix of partial wave amplitudes is given by
\begin{equation}
\left( \mathbf{a_0} \right)_{i,f}\, =\, \frac{1}{16 \pi s}\, \int_{-s}^0 dt\; \mathcal{M}_{i \to f}(s,t)\, ,
\label{eq:unit_amplitude}
\end{equation}
and the $a_j^{(0)}$ are the eigenvalues of $\mathbf{a_0}$. Again, these conditions are relevant to constrain the scalar potential parameters $\lambda_i$.

\subsection{Electroweak constraints}
\label{sec:EWPO}

The electroweak precision observables (EWPOs) measured at LEP and SLC are also included in the analysis. Since the choice of nuisance parameters does not affect significantly the results, we employ best-fit fixed values for the SM inputs $M_Z$, $m_t$, $\alpha_s$ and $\Delta \alpha^{(5)}(M_Z)$. The study of the oblique parameters $S$, $T$ and $U$ \cite{Peskin:1990zt, Peskin:1991sw, Haber:2010bw}, which are very sensitive to the scalar mass splittings, is not enough to disentangle the A2HDM contributions because of the presence of additional $Z$-vertex corrections \cite{Hollik:1986gg,Hollik:1987fg}. The most relevant ones are the quantum corrections to $\Gamma(Z\to b\bar b)$, which are enhanced by the large value of the top-quark mass~\cite{Bernabeu:1987me,Bernabeu:1990ws,Haber:1999zh}. 
We take this into account by making first a combined fit of EWPOs, excluding the ratio
$R_b\equiv\Gamma(Z\to b\bar b)/\Gamma(Z\to\mathrm{hadrons})$~\cite{Degrassi:2010ne,Haber:1999zh}. The updated results of this fit can be seen in Table~\ref{table:STU}, which updates the analysis of Ref.~\cite{deBlas:2016ojx}. These allowed ranges for the oblique parameters are then used, together with the measured value of $R_b$, to constrain the A2HDM.

\begin{table}[htb]
    \centering
    \begin{tabular}{|c|c| p{1cm}p{1cm}p{1cm} | }
    \hline
      &  \textbf{Result} & \multicolumn{3}{|c|}{ \textbf{Correlation Matrix}} \\
      \hline
      $S$ &  $ 0.093 \pm 0.101$ & 1.00 & 0.86  & -0.54 \\
      $T$ &  $0.111 \pm 0.116 $ & 0.86 & 1.00 & -0.83 \\
      $U$ &  $-0.016 \pm 0.088$ & -0.54 & -0.83 & 1.00 \\
      \hline
    \end{tabular}
    \caption{Results for the fit of the oblique parameters $S, T$ and $U$ without $R_b$.}
    \label{table:STU}
\end{table}

\subsection{Higgs constraints}
\label{sec:Higgsconstraints}

The Higgs signal strengths are defined as the ratio of the production cross section $\sigma_i$ times the branching ratio $\mathcal{B}_{f}$, over the SM prediction, for a given production channel ($i = $ ggF, VBF, VH, ttH) and decay mode ($f = \bar{b}b, \, \gamma \gamma, \, \mu^+ \mu^-,\,  \tau^+ \tau^-, \, WW,\,  Z\gamma, ZZ$),

\begin{equation}
\mu_i^f\, =\, \frac{\left(\sigma_i \cdot \mathcal{B}_f \right)_{\text{A2HDM}}}{\left(\sigma_i \cdot \mathcal{B}_f \right)_{\text{SM}}}\, =\, \frac{r_i \cdot r_f }{\sum_{f'} r_{f'} \cdot \mathcal{B}_{SM}(h \to f') } \, ,
\label{eq:strengths}
\end{equation}
where $r_{i,f}$ are the ratios of the production cross section $\sigma_i$ and decay width $\Gamma_f$, respectively, with respect to their SM predictions. 

The signal strengths are calculated in the narrow-width approximation and depend on the alignment parameters, the mixing angle $\tilde{\alpha}$ and the scalar potential parameters.
The input used contains LHC data (Run I and II) from the ATLAS and CMS collaborations, and data collected by D0 and CDF at the Tevatron. The data entering our fit are detailed in Appendix~\ref{sec:data} (Table~\ref{tab:strengths}).

Information about heavy Higgs searches of ATLAS and CMS, both at Run I and II, is summarized in Tables \ref{tab:directsearches4}, \ref{tab:directsearches1}, \ref{tab:directsearches2} and \ref{tab:directsearches3}, also in Appendix~\ref{sec:data}.
The analyses provided are quoted as 95\% upper limits, for different production and decay channels, on either $\sigma \cdot \mathcal{B}$ \ or \ $\left(\sigma \cdot \mathcal{B}\right) / \left(\sigma \cdot \mathcal{B}\right)_{\text{SM}}$, as functions of the resonance masses in the narrow width approximation.

Since low-energy constraints are not considered in this work, direct searches from LEP are not included in the fits. Upcoming direct searches from LHC can be easily added and the fits shown below can be updated.

\subsection{Flavour constraints}
\label{sec:Flavourconstraints}

\begin{figure}[!ht]
    \centering
 \includegraphics[scale=0.7]{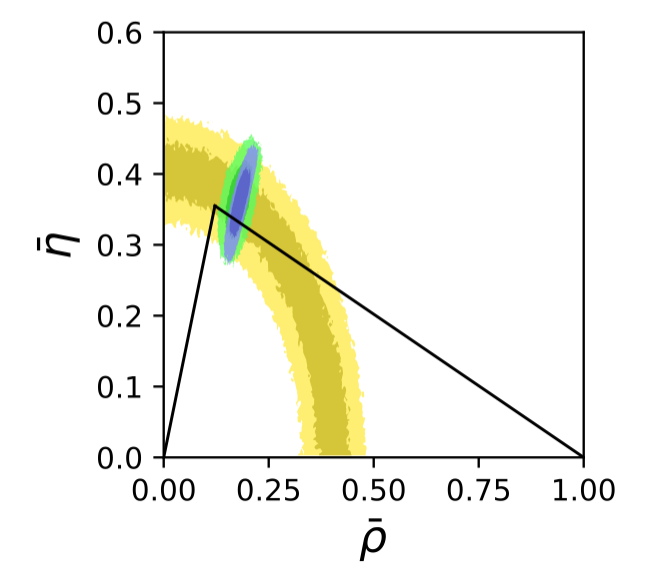}
    \qquad
    \begin{tabular}[b]{cc}\hline
      \textbf{Input}  & \textbf{Value} \\ \hline
      $\lambda$ & $0.2256 \pm 0.0009$ \\
      $A$ & $0.829  \pm     0.017$  \\
     $\bar{\rho}$ & $  0.182 \pm  0.016$ \\
     $\bar{\eta}$ & $ 0.360 \pm 0.035$  \\
    \hline
     $\rho_{\lambda, A}$  & $ -0.39$ \\
     $\rho_{\bar{\rho}, \bar{\eta}}$  & $\phantom{-}0.82$ \\ \hline
     & \\
     & \\
     & \\
    \end{tabular}
\captionlistentry[table]{entry for table}
 \captionsetup{labelformat=andtable}
  \label{fig:CKMfit}   
    \caption{Results of the CKM fit. Fitting only tree-level observables, gives the allowed regions in yellow. The green regions include $\Delta M_{B_d}$ and $\Delta M_{B_s}$ as separate observables, and the blue regions the ratio $V_{td}/V_{ts}$ from HFLAV \cite{Amhis:2019ckw} used for our fits. Darker and light colours correspond to 68\% and 95.5\%  probability, respectively.  The apex of the black triangle is   the best fit point from the SM CKM fit to all observables \cite{Tanabashi:2018oca}. The numerical values corresponding to the blue region are given in Table 2. $\rho_{i,j}$ denote the correlations between the two parameters $i$ and $j$.}
  \end{figure}

Since most of the standard CKM fits assume the SM and this would not be consistent with the study of NP, the choice of the CKM parameters is subtle. To avoid inconsistencies, a fit to the CKM entries is performed. $V_{ud}$ is extracted from superallowed $(0^+ \to 0^+)$ nuclear $\beta$ decays \cite{Hardy:2014qxa}. Given the very small value of $V_{ub}$, this fixes $V_{us}\approx\lambda$ through CKM unitarity.\footnote{Owing to a recent recalculation of the nucleus-independent radiative corrections to superallowed nuclear $\beta$ decays \cite{Seng:2018yzq,Hardy:2020qwl}, the PDG 2020 \cite{Zyla:2020zbs} value of $V_{ud}$ is about $2\sigma$ smaller than the one quoted in the PDG 2018 compilation \cite{Tanabashi:2018oca}, which implies a large ($> 3\sigma$) violation of unitarity in the first row of the CKM  matrix. 
If confirmed, this violation could not be accommodated within the A2HDM where the unitarity of the CKM matrix is exact. Improved estimates of radiative corrections are needed to resolve this issue. Meanwhile, we have adopted the PDG 2018 value of $V_{ud}$ that fits better with the kaon determination of $V_{us}$.}
$|V_{ub}|$ and $V_{cb}\approx A \lambda^2$ are obtained by combining exclusive and inclusive measurements of $b \to u \bar{\nu}_{\ell} \ell$ and $b \to c \bar{\nu}_{\ell} \ell$ transitions \cite{Amhis:2019ckw}. 
Finally, the apex $(\bar{\rho},\bar{\eta})$ of the unitarity triangle is determined with the additional information of the ratio $|V_{td}/V_{ts}|$, extracted from $\Delta M_{B_s}/\Delta M_{B_d}$ \cite{Amhis:2019ckw} that is not sensitive to charged scalar contributions \cite{Jung:2010ik}.
The CKM inputs obtained in this way and later used in our global fits are summarized in \tp{Table~\ref{fig:CKMfit} and Fig.~\hyperref[fig:CKMfit]{7.1}}.\footnote{This and the rest of the plots of the paper have been generated using \texttt{Matplotlib}  \cite{Hunter:2007ouj}.}

Charged-scalar exchanges contribute to neutral meson mixing through one-loop box diagrams \cite{Jung:2010ik, Chang:2015rva, Cho:2017jym}. The corrections induced by virtual top quarks are quite sizeable, specially for $\Delta M_{B_{s,d}}$ and $\varepsilon_K$, and provide strong constraints on $|\varsigma_u|$ (also $R_b$) as function of $M_{H^\pm}$.
The weak radiative decay $B \to X_s \gamma$ \cite{Jung:2010ik, Jung:2010ab, Jung:2012vu, Misiak:2006ab, Hermann:2012fc, Bobeth:1999ww, Misiak:2006zs} gives also important correlated constraints on $\varsigma_u$ and $\varsigma_d$, specially for large values of $\abs{\varsigma_u \varsigma_d}$. The region $\varsigma_u\varsigma_d <0$ is actually excluded, except for very small values of the alignment parameters \cite{Jung:2010ik}.
NNLO corrections \cite{Misiak:2015xwa,Misiak:2017woa,Misiak:2020vlo} are quite relevant for this observable and should be taken into account.

A complete one-loop calculation within the A2HDM of the decay $B_s \to \mu^+ \mu^{-}$ was performed in \cite{Li:2014fea,Arnan:2017lxi}. 
This observable involves the $b\to s\mu^+ \mu^{-}$ four-fermion operators $\mathcal{O}_{10}$, $\mathcal{O}_S$ and $\mathcal{O}_{P}$. The decay amplitude receives contributions from both charged and neutral scalars, and provides complementary information on the alignment parameters $\varsigma_{u,d,\ell}$ and the scalar masses. It also includes small contributions from higher-order FCNC local interactions, needed to reabsorb UV divergences, which are assumed to be negligible here. A study of these effects can be found in \cite{Penuelas:2017ikk}. Our fits include the constraints from $B_s \to \mu^+ \mu^-$, $B \to X_s \gamma$ and  $\Delta M_{B_{s,d}}$ (and $R_b$, which is discussed together with electroweak precision observables). 

Finally, the muon anomalous magnetic moment, calculated within the A2HDM in Refs.~\cite{Ilisie:2015tra,Cherchiglia:2017uwv}, is of interest because it shows a deviation with respect to the SM that, if confirmed, would strongly constrain the leptonic alignment parameter $\varsigma_{\ell}$. Its implications will be discussed in Section~\ref{sec:FitResults_Light}.

\section{Results: \textit{light scenario} }
\label{sec:FitResults_Light}

In this section, we present our main results, obtained in the  {\it light scenario} which assumes that the observed SM Higgs is the lightest CP-even  scalar of the model. The complementary possibility (the observed scalar is the heaviest) will be briefly discussed in Section~\ref{sec:FitResults_Heavy}. 
We analyse first the separate implications of the different types of constraints, before combining all of them into a final global fit to the data.

\subsection{Theoretical constraints}
\label{sec:TheoryResults}

\begin{figure}[h!]
    \centering
    \includegraphics[scale=0.35]{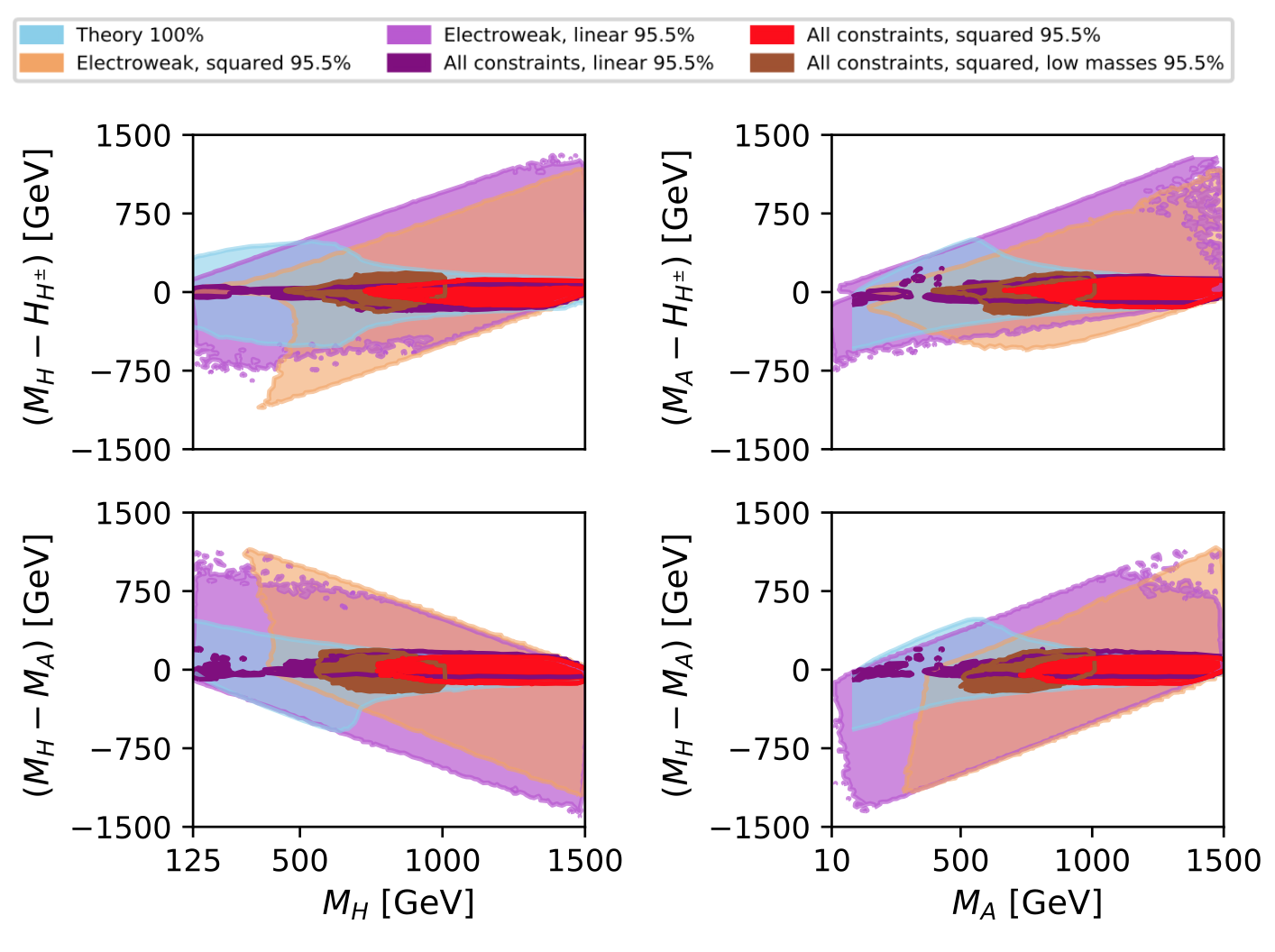}
    \includegraphics[scale=0.33]{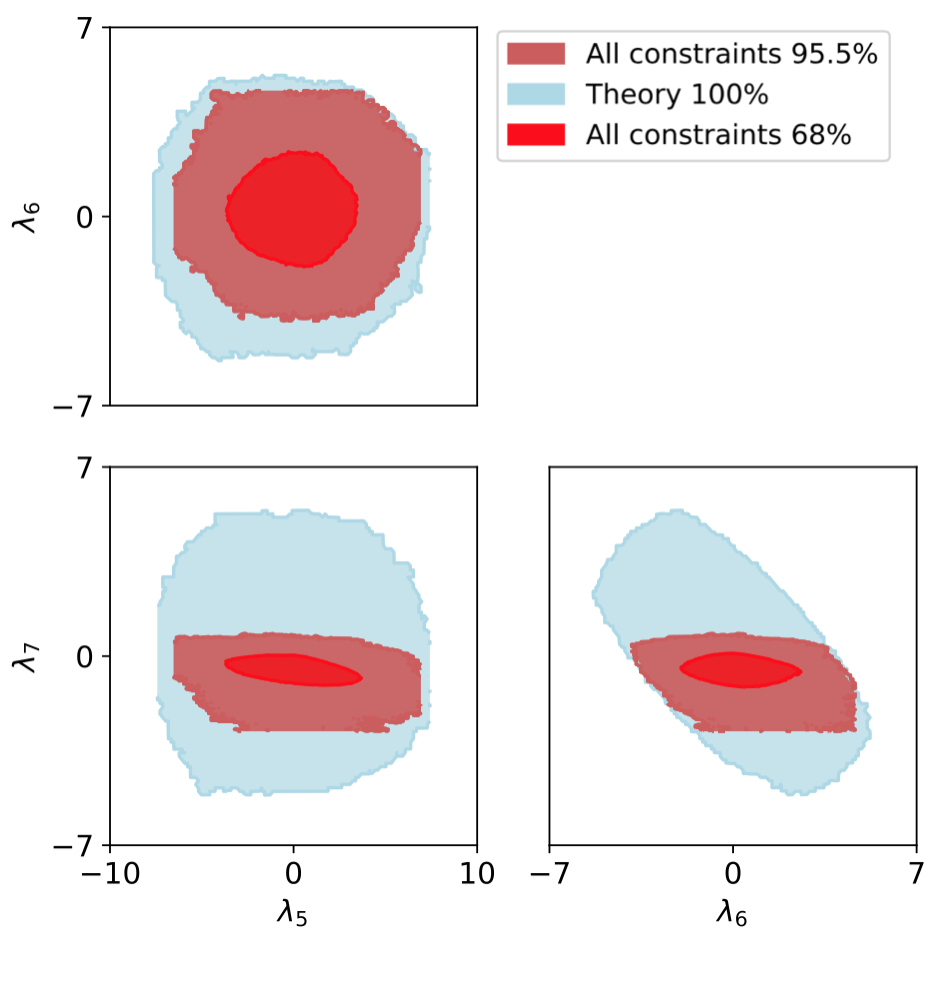}
    \caption{\textbf{Left panel:} Allowed regions for the scalar mass splittings coming from theoretical constraints at 100\% probability (blue), from EWPOs at 95.5\% probability (in orange, squared mass priors and  in light purple, linear priors), and combining all constraints at 95.5\% probability (linear mass priors in purple, squared mass priors in red and squared mass priors with lower masses in brown). The ``All constraints'' contains only the right-sign branch discussed in Section~\ref{sec:GlobalFit}. \textbf{Right panel:} Two-dimensional bounds on the $\lambda_5, \lambda_6$ and $\lambda_7$ parameters of the potential
    resulting from imposing theoretical constraints  (blue, 100\% probability), and considering all constraints (in dark red 95.5\% probability, in red 68\% probability). }
    \label{fig:Mass_Massdiff_Lambda}
\end{figure}

\begin{figure}[h!]
    \centering
    \includegraphics[scale=0.5]{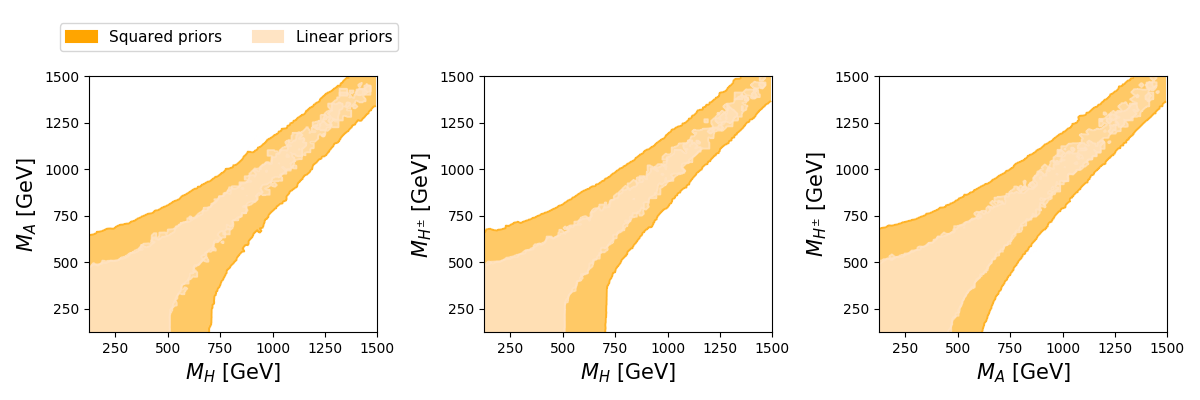}
    \caption{Theoretical constraints (100\% probability) obtained with squared mass priors (dark orange) and linear mass priors (light orange).  }
    \label{fig:Theo_priors}
\end{figure}

Perturbative unitarity and positivity of the scalar potential set strong limits on the scalar masses and the quartic parameters $\lambda_i$. The mass differences among $H$, $A$ and $H^\pm$ are strongly constrained, as shown by the allowed blue regions in the left panel of Fig.~\ref{fig:Mass_Massdiff_Lambda} (electroweak and combined constraints, also present in the plots, will be commented later in Sections~\ref{sec:ElectroweakResults} and~\ref{sec:GlobalFit}):
\begin{equation}
    \abs{M_i - M_j} \leq 600 \text{ GeV,} \qquad\qquad i,j = H, A, H^{\pm}, \qquad \text{(squared mass priors)}.
    \label{eq:massdiff}
\end{equation}
%
There is a clear correlation between the masses of any two scalar particles: a large mass for one scalar implies that the other scalar mass is also restricted to be large. This effect is stronger for higher values of the scalar masses. The allowed mass splittings decrease as the average mass scale increases. This is easily understood from the scalar mass relations in Eqs.~(\ref{eq:MAHpm}) and (\ref{eq:MhH}), since $M_{H^\pm}$, $M_H$ and $M_A$ become degenerate in the limit  $\mu_2\gg \lambda_i v^2$. 
The impact of the assumed mass priors is further studied in Fig.~\ref{fig:Theo_priors}.
Both linear (light orange) and squared (dark orange) options 
give rise to allowed regions with similar shapes, although they are larger for the squared priors.

The theoretical constraints also restrict the allowed ranges of the scalar quartic couplings.
Two-dimensional plots in the space ($\lambda_5, \lambda_6, \lambda_7$) are shown in the right panel of Fig.~\ref{fig:Mass_Massdiff_Lambda}, which displays the correlations among these three parameters of the scalar potential.
The positivity relations imply bounds on $|\lambda_5|$ and $|\lambda_6+\lambda_7|$ because these two quantities
induce negative contributions to the two last conditions in Eq.~\eqref{eq:theoLambda}.  This implies an anti-correlation between $\lambda_6$ and $\lambda_7$, which is clearly manifest by the allowed blue area in the $\lambda_6-\lambda_7$ plane.
The blue regions in Fig.~\ref{fig:Mass_Massdiff_Lambda} satisfy the bounds derived in previous works \cite{Branco:2011iw}.

\subsection{Electroweak constraints}
\label{sec:ElectroweakResults}

\begin{figure}[t!]
    \centering
    \includegraphics[scale=0.55]{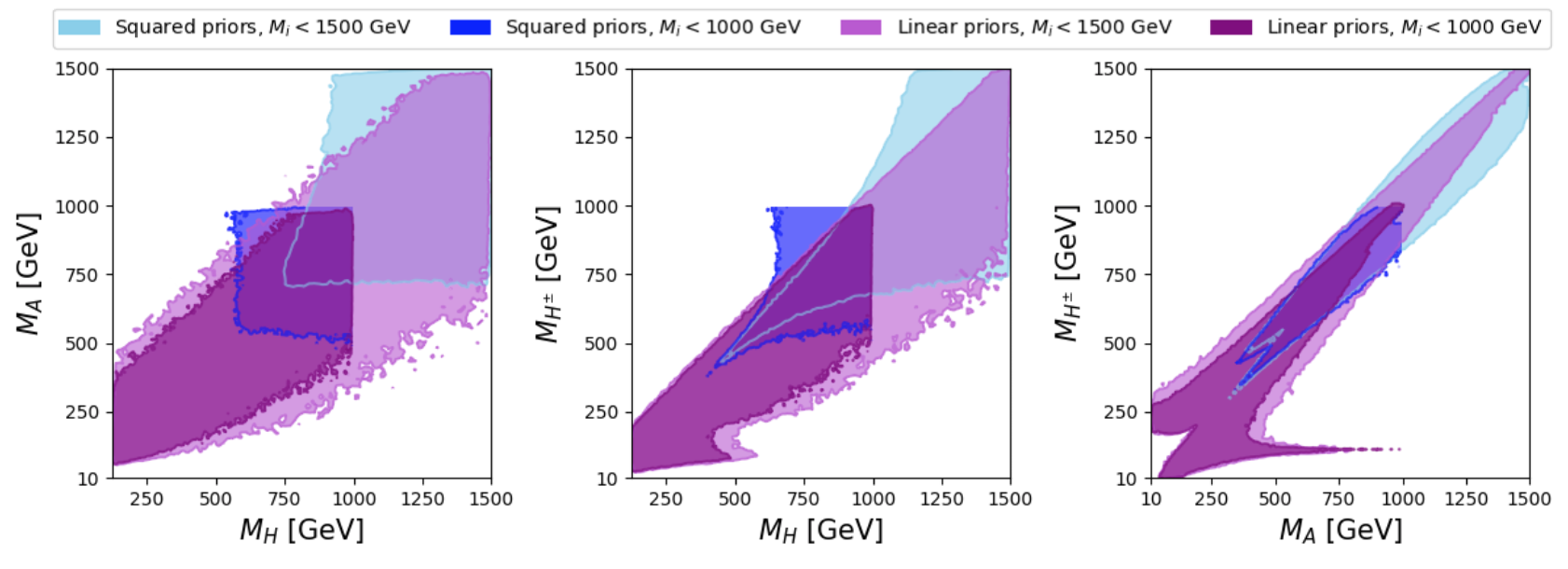}
    \caption{Allowed mass ranges from EWPOs at a 68\% probability. Light and dark blue (purple) correspond to squared (linear) mass priors with $M_{H^{\pm},A} \in [10, 1500]$ GeV and $M_H \in [125, 1500]$ GeV (lighter regions) and  with $M_{H^{\pm},A} \in [10, 1000]$ GeV and $M_H \in [125, 1000]$ GeV (darker regions).}
    \label{fig:Mass_priors}
\end{figure}

The EWPOs restrict the individual masses of the scalar particles in the low-mass range, and are very useful to constrain their mass splittings. The oblique parameters are very sensitive to the scalar mass differences, which results in strong limits for the masses. This can be clearly observed in Figs.~\ref{fig:Mass_priors} and \ref{fig:Mass_Mass}. The information from EWPOs complements in a very useful way the theoretical constraints discussed before. 

The allowed regions obtained from EWPOs present a strong dependence on the mass priors, as can be seen in Fig.~\ref{fig:Mass_priors}, which displays the limits resulting from different choices of mass priors. Independently of the priors, large values for the masses and small  splittings are favoured.  The light and dark blue regions show a very strong dependence on the assumed ranges for mass-squared priors. If the scalar masses are varied until 1500 GeV, masses below approximately 750 GeV are not allowed at a 68\% probability, while if a lower range below 1000 GeV is adopted, scalar masses of 500 GeV are allowed at the same probability. The same tendency is observed if lower mass regions are chosen. 
If the fit is repeated with linear mass priors (purple regions), masses as low as 10~GeV are allowed for the charged and CP-odd neutral scalars, at a 68\% probability. In this case, the dependence on the input mass ranges is also weaker. The allowed regions for linear mass priors up to 1000~GeV (1500~GeV) are indicated in dark (light) purple colour. 

The orange and light purple regions in the left panel of Fig.~\ref{fig:Mass_Massdiff_Lambda} display the constraints from EWPOs on the scalar mass splittings with squared and linear priors, respectively. These allowed regions have been obtained varying the mass priors in their full range up to 1500~GeV.

\subsection{Higgs constraints}
\label{sec:HiggsResults}

\subsubsection{Higgs signal strengths}

Since the measured Higgs signal strengths are consistent with the SM, within their current uncertainties, the gauge and Yukawa couplings of the SM-like Higgs boson should be close to the SM limit. In particular, the measured data on the $WW^*$ and $ZZ^*$ decay modes imply that $\cos{\tilde\alpha}$ cannot deviate much from one and, therefore, $\tilde\alpha$ should be small.
A similar comment applies to the $H f\bar f$ interactions. However, most Higgs observables are not sensitive to the signs of the Yukawa couplings and, therefore, the LHC data only require the modulus of $|y_f^h|-1$ to be smaller than about 0.1-0.2. This gives two different types of solutions for the Yukawa couplings:
there will be a broad range of allowed values of $\varsigma_f$ with $\tilde{\alpha} \approx 0$, corresponding to $y_f^h~\approx~1$, and another region with somewhat larger values of the mixing angle corresponding to  $y_f^h~\approx~-1$.  

\begin{figure}[t]
    \centering
    \includegraphics[scale=0.5]{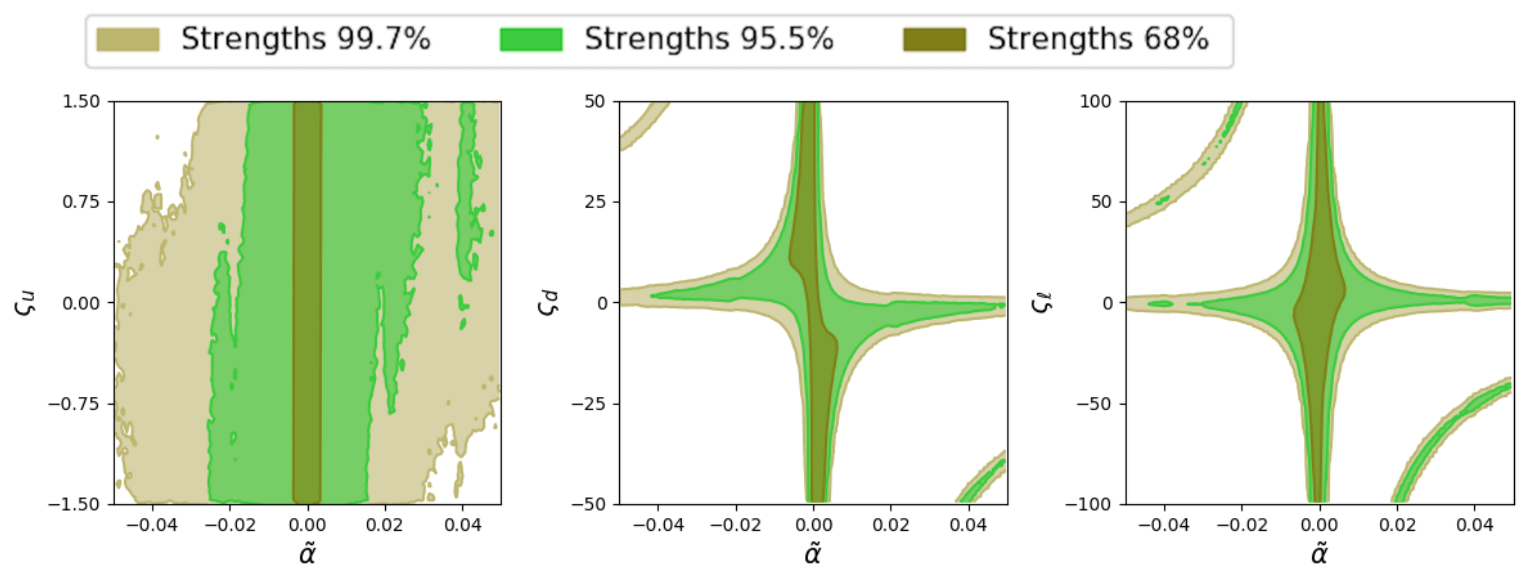}
    \caption{Constraints on the planes $\tilde{\alpha} - \varsigma_f$ from the Higgs signal strengths at a 68\% (dark green), 95.5\% (light green) and 99.7\% (olive green) probability.}
    \label{fig:alpha_couplings_S}
\end{figure}

For small values of $\tilde{\alpha}$, Eq.~\eqref{eq:y_coup} gives 
$ y_{f}^h = 1 + \tilde{\alpha} \, \varsigma_{f} + \mathcal{O}(\tilde{\alpha}^2)$ (assuming $h$ to be the lightest CP-even neutral scalar), so that the Yukawa coupling is close to -1 for $\tilde{\alpha} \, \varsigma_{f} \approx -2$.
This effect can be observed in the allowed $(\tilde{\alpha}, \varsigma_f)$ regions of Fig.~\ref{fig:alpha_couplings_S}, for the down-quark and lepton alignment parameters, where separate $\tilde{\alpha} \, \varsigma_{f} \ll 1$ and $\tilde{\alpha} \, \varsigma_{f} \approx -2$ solutions are clearly visible.
The up-quark Yukawa sign ambiguity gets broken by the two-photon decay amplitude of the Higgs that involves one-loop contributions from virtual $W^\pm$, $t$ and $H^\pm$. Assuming that the charged-scalar correction is small, the measured $H\to\gamma\gamma$ signal strength determines the relative sign between $y_u^h$ and $g_{hWW}$ to be positive. Therefore, only the region $\tilde{\alpha} \, \varsigma_{u} \ll 1$ is allowed in this case.

In the following we will distinguish among the two different possibilities: the ``right-sign'' solution, corresponding to $y_{d, \ell}^h~\approx~1$ and the ``wrong-sign'' one corresponding to $y_{d, \ell}^h~\approx~-1$.  The  former was previously analysed in the A2HDM~\cite{Celis:2013ixa} and, more recently, in the particular case of $\mathcal{Z}_2$ symmetric models \cite{Chowdhury:2017aav}.
For the  ``right-sign'' solution we find that the value of $\tilde{\alpha}$ is strongly constrained (radian units): 
\begin{eqnarray}
\nonumber
\abs{\tilde{\alpha}} &\leq& 0.003
\qquad\qquad (68\% \;\mathrm{probability}),
\\ 
\abs{\tilde{\alpha}} &\leq& 0.023
\qquad\qquad  (95.5\% \;\mathrm{probability)}.
\label{eq:alpha_S}
\end{eqnarray}

\subsubsection{Direct searches}

The negative results from direct searches restrict the masses of the scalar particles.
In order to access to the information that these observables provide, we first calculate the theoretical production cross section times branching ratio $\sigma \cdot \mathcal{B}$ in the A2HDM. We consider then the ratio
$R \equiv (\sigma \cdot \mathcal{B})^{\text{theo}}/(\sigma \cdot \mathcal{B})^{\text{obs}}$
between the theoretical value and the observed limit, to which we assign a Gaussian likelihood with zero central value, which is in agreement with the null results obtained so far in the searches of heavy scalars. The corresponding standard deviation of the likelihood is adjusted in a way that the value  $R=1$ can be excluded with a probability of the 95\%. The production cross sections and branching ratios for the other scalar particles are calculated in a similar way to the SM Higgs, taking into account the kinematically allowed region and the CP quantum number of the particle.

In general, the data from direct searches favour heavier scalars and help us to restrict lower masses. However, since there are less experimental searches in the low-mass range, one gets less restrictive constraints for masses below $100$~GeV. The constraints available so far seem to indicate that low masses are still allowed, so information from direct searches in that region would be crucial to understand the phenomenology at low masses.

\subsection{Flavour constraints}
\label{sec:FlavourResuls}

\begin{figure}[t!]
    \centering
\includegraphics[scale=0.5]{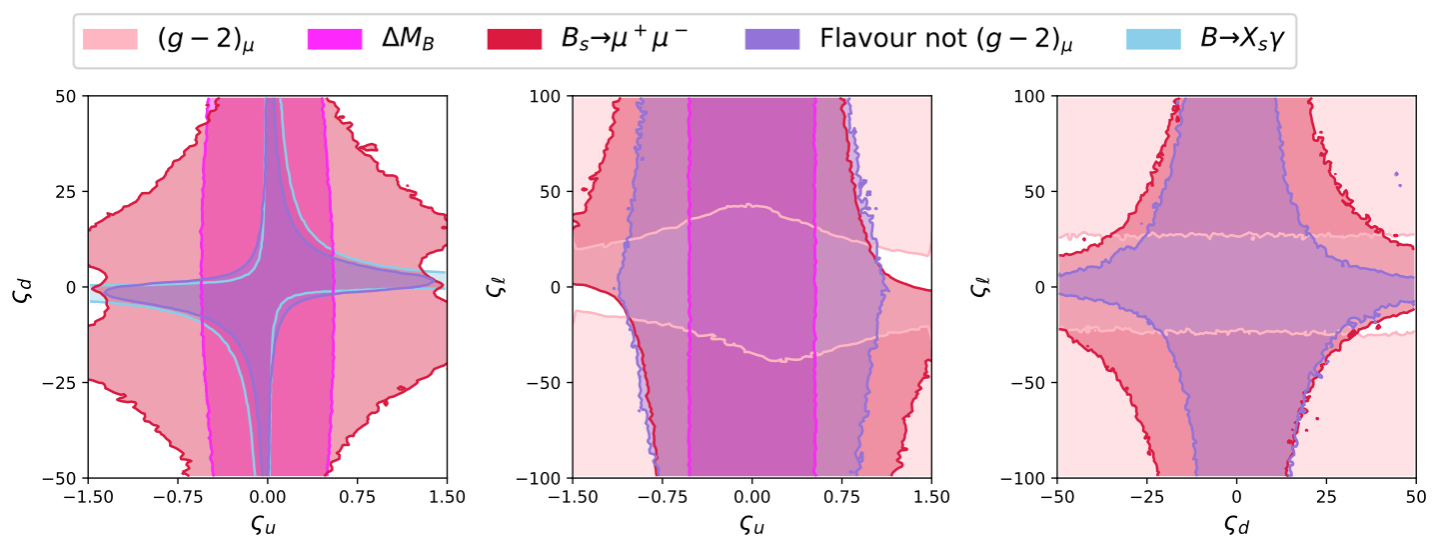}
    \caption{Constraints at the 95.5\% probability on the alignment parameters from $(g-2)_{\mu}$ (light pink), $B_s \to \mu^+ \mu^-$ (red), meson mixing (dark pink), $B \to X_s \gamma$ (blue) and all flavour observables except $(g-2)_{\mu}$ (purple). For clarity observables that do not give relevant constraints in a given plane are omitted from the plot.}
    \label{fig:Coup_flavour_2}
\end{figure}

Flavour observables are useful to constrain the Yukawa alignment parameters $\varsigma_f$. 
Fig.~\ref{fig:Coup_flavour_2} displays the allowed (95\% probability) two-dimensional regions in the three-dimensional $\varsigma_f$ space from independent analyses of the most relevant flavour measurements: $B^0$ mass mixing (dark pink), $B_s \to \mu^+ \mu^-$ (red),  $B \to X_s \gamma$ (blue) and $(g-2)_{\mu}$ (light pink). For clarity the observables that do not give relevant constraints in a given plane are omitted from the corresponding plot.

As already known from previous works \cite{Ilisie:2015tra,Cherchiglia:2017uwv,Wang:2018hnw}, the $(g-2)_{\mu}$ anomaly requires sizeable NP contributions, which translates into  non-zero values for $\varsigma_{\ell}$ that are rather large, while it is insensitive to $\varsigma_d$. This can be clearly seen in the central and right panels of Fig.~\ref{fig:Coup_flavour_2} where the light-pink regions exclude values of $|\varsigma_{\ell}|$ below 10-20. The precise size of the discrepancy with the SM expectation relies, however, in a phenomenological evaluation of the hadronic contribution to the photon vacuum polarization (and a smaller light-by-light hadronic correction), involving a very subtle combination of different experimental data sets, which not always are in good agreement \cite{Aoyama:2020ynm}.
Thus, the statistical relevance of the $(g-2)_{\mu}$ anomaly could be magnified by
underestimated uncertainties. The most recent lattice calculations seem in fact to suggest that the SM prediction could be much closer to the current  $(g-2)_{\mu}$ measurement \cite{Pich:2020gzz}.
While waiting for a possible confirmation (or not) of this intriguing anomaly, and in order not to bias the results, we will not include $(g-2)_{\mu}$ in the global fit. We will discuss later whether the large values of $|\varsigma_{\ell}|$ currently required to accommodate $(g-2)_{\mu}$ are compatible with the parameter ranges emerging from a global fit to the other observables.

For similar reasons, the recent flavour anomalies observed in $b\to c \tau\nu$ and $b\to s\mu^+\mu^-$ transitions \cite{Pich:2019pzg} will not be included either in our global fit. A recent model-independent analysis of the $b\to c \tau\nu$ anomaly has been already given in \cite{Murgui:2019czp,Mandal:2020htr}, where references to previous works can be found.
If confirmed, these anomalies would provide clear evidence of NP with non-universal lepton couplings. 

The remaining flavour observables do not show significant deviations from the SM and, therefore, lead to allowed regions in Fig.~\ref{fig:Coup_flavour_2} with smaller values of the alignment parameters, including the null SM solution. $B_s \to \mu^+ \mu^-$ is the only observable that constrains the leptonic couplings, excluding large values for $\abs{\varsigma_{u,d}\varsigma_{\ell}}$ as indicated by the magenta areas in the figure.
A similar restriction on the product $|\varsigma_{u} \varsigma_d|$ can be appreciated in the left panel.

The mass differences between the neutral $B^0$ eigenstates, $\Delta M_{B_d}$ and $\Delta M_{B_s}$,
are dominated by virtual top-quark contributions. This results in a strong upper limit on $|\varsigma_u|$, which corresponds to the vertical pink bands in the left and central panels of  
Fig.~\ref{fig:Coup_flavour_2}. Similar limits emerge from $\varepsilon_K$ and $R_b$ \cite{Jung:2010ik}. However, it can be seen that values of $|\varsigma_u|$ excluded at a 95.5\% probability from $\Delta M_B$ are no-longer excluded at the same probability when all flavour observables are combined.
A stronger constraint on the $\varsigma_u - \varsigma_d$ plane can be derived from the radiative decay $b \to s \gamma $. The blue region in the left panel of the figure shows the strong correlation between the two quark alignment parameters, which prevents them to be large simultaneously.

\subsection{Global fit}
\label{sec:GlobalFit}

After discussing the separate effect of each type of observables, let us analyse the limits emerging from the global fit to all experimental and theoretical inputs.

\begin{figure}[ht!]
    \centering
    \includegraphics[scale=0.55]{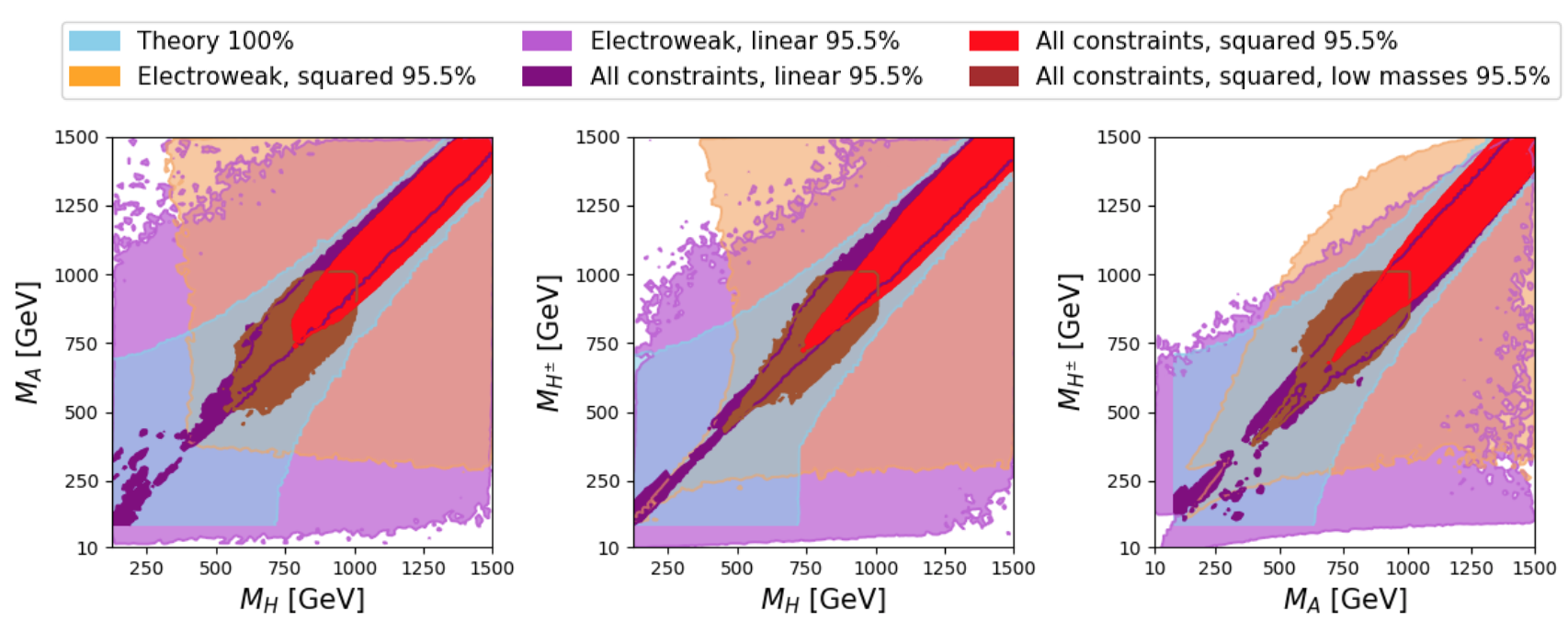}
    \caption{Allowed regions on the planes $M_H$--$M_A$ (left), $M_H$--$M_{H^{\pm}}$ (middle) and $M_A$--$M_{H^{\pm}}$ (right) from theoretical constraints at 100\% probability (blue), from EWPOs at 95.5\% probability (in orange, squared mass priors and  in light purple, linear priors), and combining all constraints at 95.5\% probability (linear mass priors in purple, squared mass priors in red and squared mass priors with lower masses in brown). The ``All constraints'' regions contain only the right-sign branch solution to the Higgs signal strengths discussed in Section~\ref{sec:HiggsResults}.}
    \label{fig:Mass_Mass}
\end{figure}

The combined constraints on the scalar masses and mass differences are shown in the left panel of Fig.~\ref{fig:Mass_Massdiff_Lambda} and in Fig.~\ref{fig:Mass_Mass}.
From these plots, it can be seen that theoretical and EWPOs constraints are complementary and by combining them with the remaining observables, light values for the masses are disfavoured. As for the electroweak constraints, there is a clear dependence on the mass priors. The global fits adopting squared mass priors with $M_i^2 \leq 1500^2~\text{GeV}^2$  and $M_i^2 \leq 1000^2~\text{GeV}^2$ are displayed  in the figures as red and brown regions, respectively. These results show that if the fitted mass regions are reduced, lower values of the scalar masses are allowed at the same probability. The global fit with linear priors is displayed in purple, for $M_i \leq 1500$~GeV, showing that lighter masses are allowed than with squared mass priors. A similar, but weaker effect is observed for the mass splittings in Fig.~\ref{fig:Mass_Massdiff_Lambda}. With the squared mass priors one can set bounds on the mass differences: 
\begin{equation}
    \abs{M_i - M_j} \leq 150 \text{ GeV,} \qquad\qquad i,j = H, A, H^{\pm}, \qquad \text{(squared mass priors).}
\end{equation}
However, the strong dependence on the mass priors indicates that these mass constraints should be taken with some care.

The right panel in Fig.~\ref{fig:Mass_Massdiff_Lambda} shows the resulting allowed regions (68\% probability in red and 95.5\% probability in dark red) for the scalar potential parameters $\lambda_i$ when all constraints are included in the global fit. The addition of the Higgs signal strengths and the direct searches restricts significantly the parameter space obtained before from theoretical observables (blue areas). This effect is specially strong for $\lambda_7$.

\begin{figure}[t]
    \centering
    \includegraphics[scale=0.55]{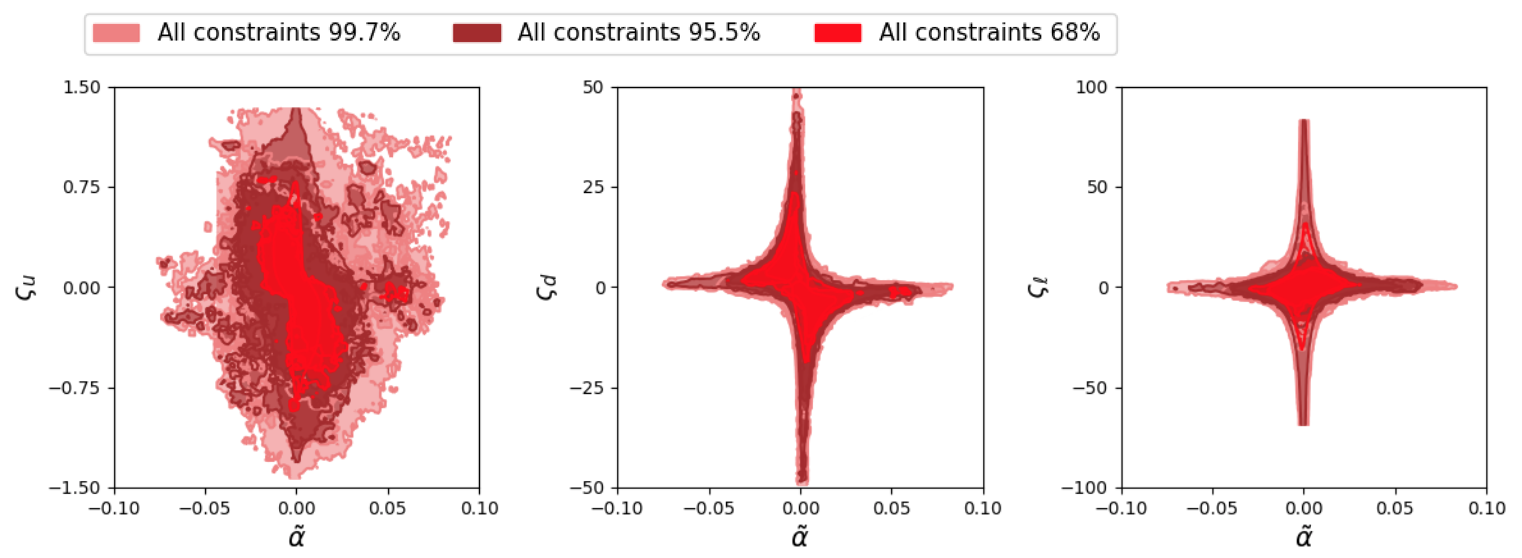}
    \caption{Constraints on the planes $\varsigma_f - \tilde{\alpha}$ from the global fit at a 68\% (dark red), 95.5\% (brown) and 99.7\% (light red) probability. Only the ``right-sign'' solution is included.}
    \label{fig:alpha_couplings_GF}
\end{figure}

Combining the information from the Higgs signal strengths with the other observables turns out to be a bit subtle because the fine-tuned ``wrong-sign'' solutions discussed in Section~\ref{sec:HiggsResults} lead to a very slow numerical convergence of the fit algorithm. To solve that, we have performed the fits shown in this section with the condition $y_{d, \ell}^h \approx 1 $. The negative branch solution will be discussed separately in Section~\ref{sec:wrongbranch}. 

For the positive branch, once we add the  rest of observables to the Higgs signal strengths, the constraints of Fig.~\ref{fig:alpha_couplings_S} get modified into the ones of Fig.~\ref{fig:alpha_couplings_GF}. The global fit gives stronger limits for the alignment parameters, as expected, but leaves a somewhat wider allowed range for the mixing angle
(radian units):
\begin{eqnarray}
 \nonumber
-0.015 &\leq \tilde{\alpha} \leq & 0.013\qquad\qquad (68\% \;\mathrm{probability)},  
\\
-0.04 &\leq \tilde{\alpha} \leq & 0.04\qquad\qquad (95.5\% \;\mathrm{probability)}.
\label{eq:alpha_GF}
\end{eqnarray}
This counterintuitive statistical effect originates from the fact that the other observables are not very sensitive to $\tilde{\alpha}$.

\begin{figure}[t]
    \centering
\includegraphics[scale=0.55]{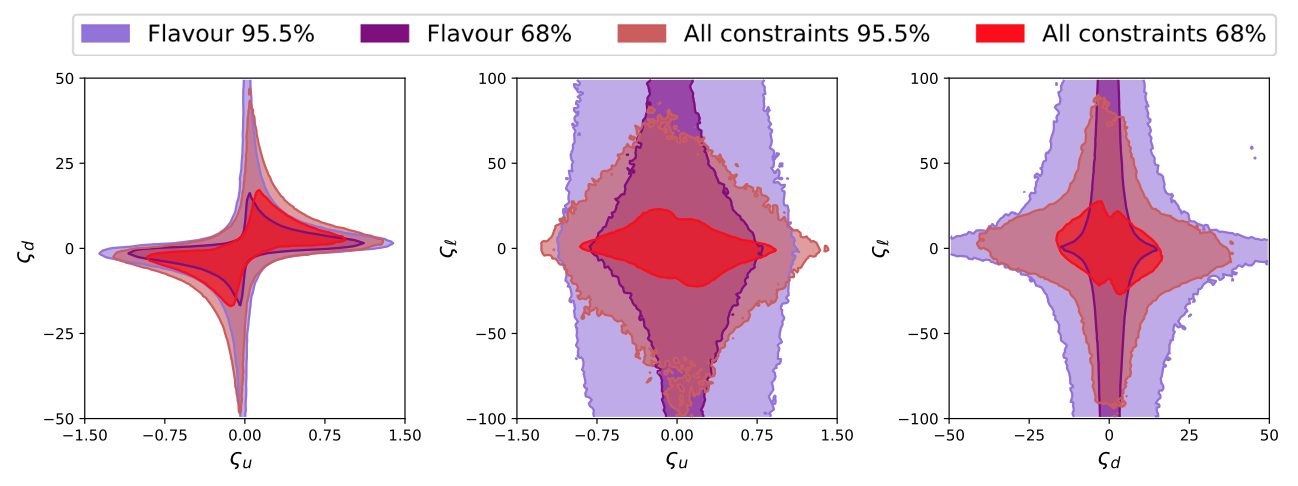}
    \caption{Constraints on the alignment parameters from a fit with only flavour observables (68\% probability, violet; 95.5\% probability, purple) and from the global fit with the positive branch solution (68\% probability, red; 95.5\% probability, brown).}
    \label{fig:Coup_all}
\end{figure}

Fig.~\ref{fig:Coup_all} compares the constraints on the alignment parameters resulting from the combination of flavour observables (not including $(g-2)_{\mu}$) with the regions allowed by the global fit at 68\% and 95.5\% probability. The limits on the down-quark and lepton couplings become stronger, once all constraints are considered. This is mainly due to the combined effect of the Higgs and flavour observables. The strong correlation between the up-quark and down-quark alignment parameters observed before remains also in the global fit, with tighter upper bounds on $|\varsigma_d|$: larger values of the up coupling require smaller values of the down coupling and vice versa. A similar but weaker effect can be observed in the $\varsigma_u - \varsigma_\ell$ and $\varsigma_d - \varsigma_{\ell}$ planes.

\subsection{``Wrong-sign'' solution}
\label{sec:wrongbranch}

\begin{figure}[ht]
    \centering
    \includegraphics[scale=0.5]{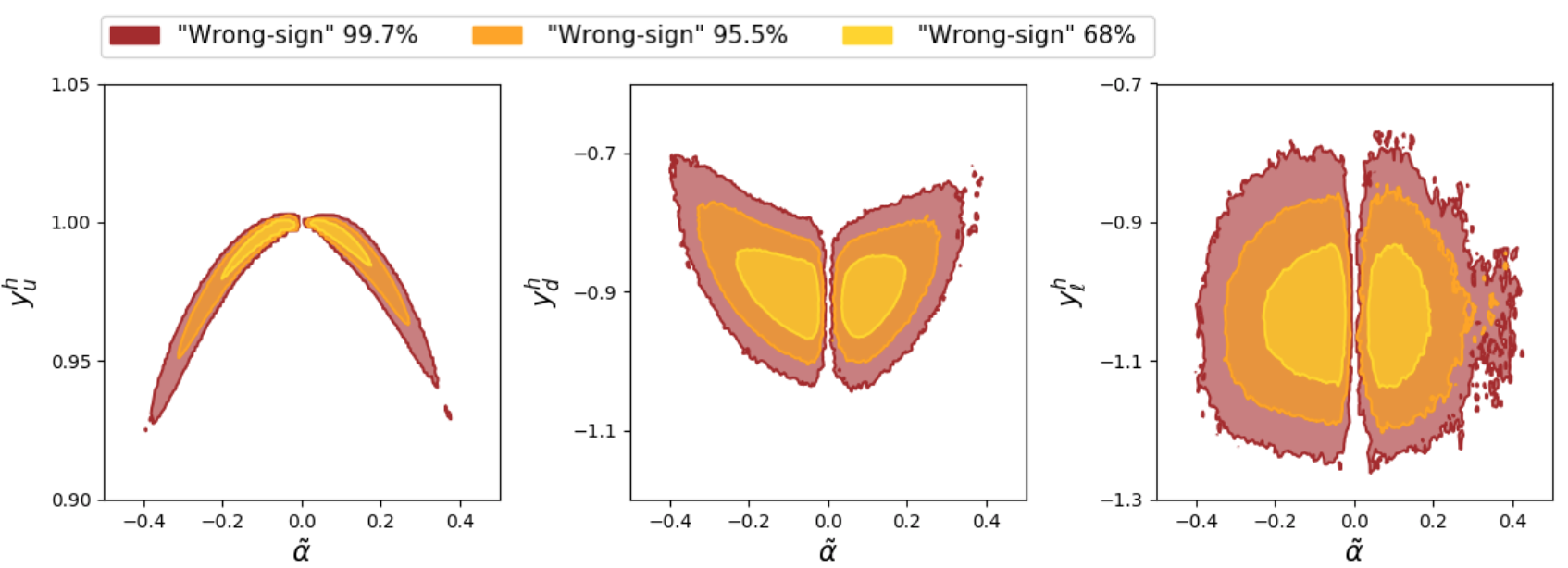}
    \caption{Two-dimensional constraints on the mixing angle $\tilde{\alpha}$ and the Yukawa couplings $y^h_f$ from a global fit with the ``wrong-sign'' solution for the Higgs signal strengths. The plots show the allowed regions at 99.7\% (brown), 95.5\% (orange) and 68\% (yellow) probability. }
    \label{fig:wrong_sign}
\end{figure}

The regions allowed by the global fit with the ``wrong-sign'' solution for the Higgs signal strengths are displayed in Fig.~\ref{fig:wrong_sign}, at different probabilities. Since these constraints have been obtained imposing the ``wrong-sign'' solution, whose probability is smaller than 100\%, the final probability would be (probability of the ``wrong-sign'' solution)$\times$(probability of Fig.~\ref{fig:wrong_sign}). The fine-tuned condition $\tilde{\alpha}\varsigma_{d,\ell}\sim -2$, emerging from $y_{d,\ell}^h \sim -1$, can be only satisfied in a small portion of the parameter space. The null value for the mixing angle is not reached, since it corresponds to $y_{f}^h = 1$ and, therefore,  belongs to the normal ``right-sign'' branch discussed in the previous subsection.

\section{Results: \textit{heavy scenario} }
\label{sec:FitResults_Heavy}

In the previous section we have described the situation in which the observed Higgs corresponds to the lightest CP-even scalar of the model. In this section we will analyse the complementary situation, \emph{i.e.} the heaviest CP-even scalar is the SM Higgs and there is an additional neutral scalar with mass below 125 GeV.

\begin{figure}[tb]
\centering
\includegraphics[scale=0.45]{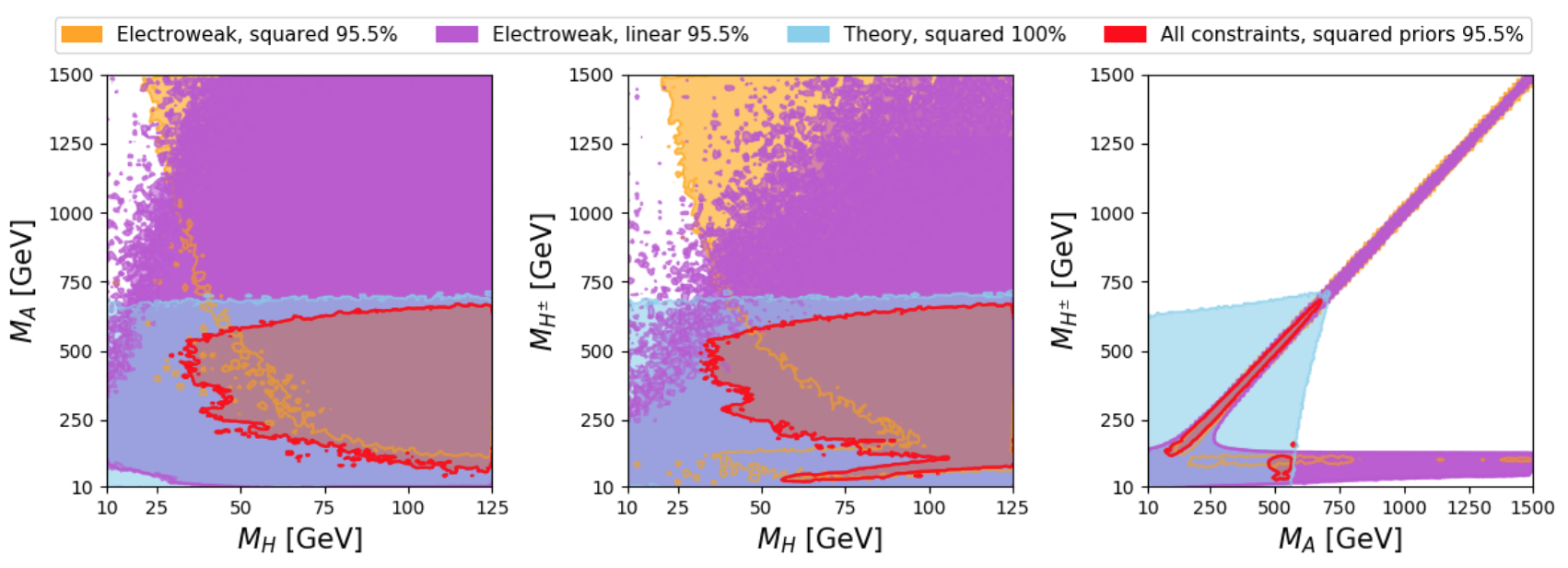}
\caption{Two-dimensional constraints on the scalar masses in the {\it heavy} scenario.
The different allowed regions correspond to theoretical constraints (blue, squared priors, 100\% probability), EWPOs (95.5\% probability; orange, squared priors; violet, linear priors) and the combined global fit
 (red, squared priors, 95.5\% probability).}
\label{fig:MassMassH}
\end{figure}

The theoretical constraints show the same tendency as for the {\it light scenario}. Since the mass of the CP-even scalar is now bounded to be light, the remaining two scalar masses cannot be heavier than $700$ GeV. This can be seen in Fig.~\ref{fig:MassMassH} for squared mass priors. Linear mass priors give very similar theoretical constraints, so they are omitted from the plot.

The constraints from EWPOs are also similar to the ones of the {\it light scenario}. Lower masses and large mass splittings are excluded when squared priors are adopted. Now $M_H$ is bounded to be smaller than 125 GeV, so the allowed regions become narrower. The worrisome difference between the results obtained with linear and squared mass priors is also present in this {\it heavy scenario}. Squared priors give very strong constraints for light masses that are no-longer found when linear priors are used.
These constraints are displayed in Fig.~\ref{fig:MassMassH}, which shows that rather low values for all scalar masses are indeed allowed by the linear-priors fit. 
In the plane $M_A - M_H^{\pm}$ the allowed regions for linear (violet) and squared (orange) priors overlap, so it not easy to distinguish them.

\begin{figure}[htb]
    \centering
    \includegraphics[scale=0.5]{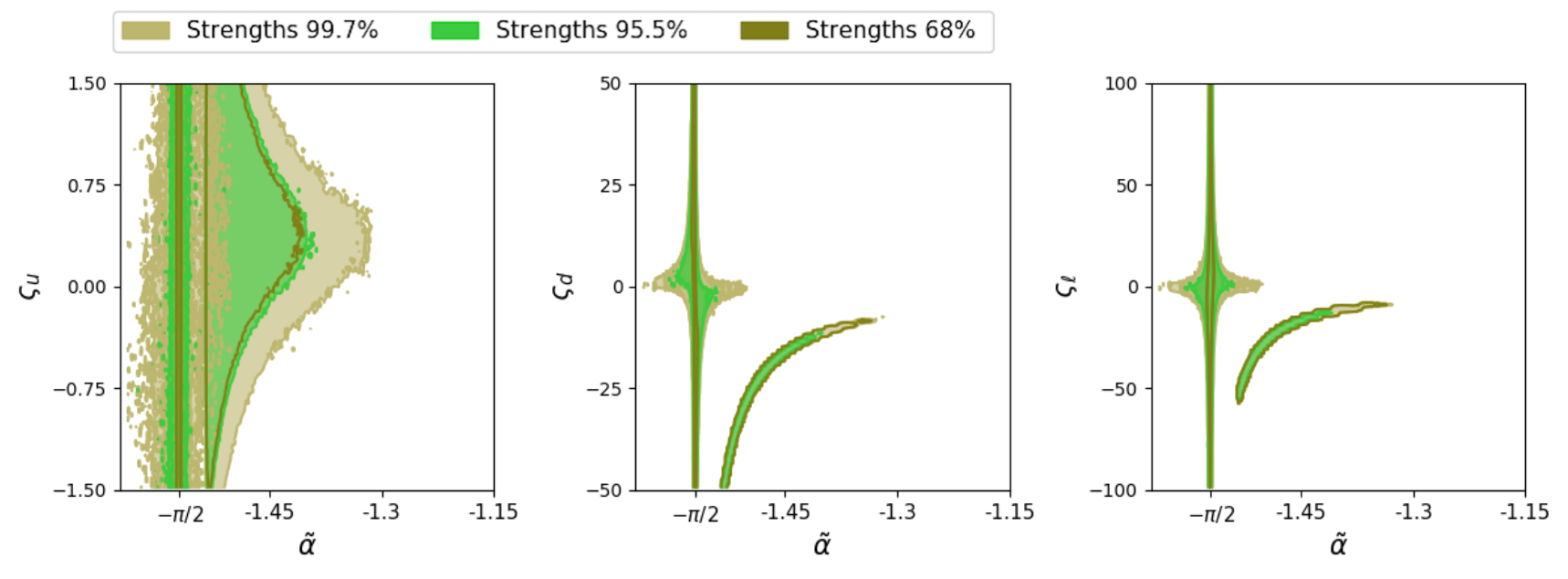}
    \caption{Constraints from Higgs signal strengths in the {\it heavy scenario}. One can distinguish the ``right-branch'' with $\tilde{\alpha} \approx - \frac{\pi}{2}$, corresponding to $y_{d,\ell}^h \approx 1$, and the  ``wrong branch'' with $(\tilde{\alpha} + \frac{\pi}{2})\,\varsigma_{d,\ell} \approx -2$ for $y^h_{d, \ell} \approx -1$.}
    \label{fig:St_heavy}
\end{figure}

The analytical expressions of the gauge and Yukawa couplings of the light and heavy CP-even neutral scalars in Eqs.~(\ref{eq:gaugecouplings}) and (\ref{eq:y_coup})
can be shifted with the change of variable $\tilde{\alpha} = \tilde{\beta} - \frac{\pi}{2}$ (notice that this brings $\tilde{\beta}$ outside our previous convention for $\tilde{\alpha}$), up to a global minus sign in the so-far unmeasured couplings of the additional neutral scalar. Therefore,  the constraints on $\tilde\beta$ from the Higgs signal strengths would be similar to the ones obtained for $\tilde\alpha$ in the {\it light scenario}. Adopting the convention that 
$g_{hVV}$ should be positive, the $hWW^*$ and $hZZ^*$ measurements imply now that $\tilde{\alpha}$ should be close to $-\frac{\pi}{2}$. The ``right-branch'' where $y^h_{d,\ell}$ have the same sign as $y^h_u$ corresponds also to the region with $\tilde{\alpha} \approx - \frac{\pi}{2}$, while the ``wrong-branch'' where $y^h_{d,\ell}$ have the opposite sign satisfies $(\tilde{\alpha} +\frac{\pi}{2})\,\varsigma_f\approx  -2$.
As for the {\it light scenario}, the up Yukawa has always the same sign as $g_{hVV}$. These allowed regions are displayed in Fig.~\ref{fig:St_heavy}. The sharp cut close to $\tilde{\alpha} = -\frac{\pi}{2}$ in the negative branch is a consequence of the correlation between $\tilde{\alpha}$ and the down coupling $\varsigma_d$. Lower values of the mixing angle would require $\varsigma_d < -50$, which is not allowed by our priors.

Finally, most flavour constraints are independent of the  neutral scalar masses, so they are identical in the {\it light} and {\it heavy} scenarios. The only relevant dependence appears in $B_s\to\mu^+\mu^-$ for large values of $\varsigma_{d,\ell}$  \cite{Li:2014fea}. The allowed regions in Fig.~\ref{fig:Coup_flavour_2} remain then also valid in the {\it heavy} scenario, except the magenta areas that get slightly distorted.

The final constraints on the scalar masses from the global fit to this {\it heavy} scenario are displayed in Fig.~\ref{fig:MassMassH}, assuming squared mass priors (red regions). Large masses for the neutral CP-odd and charged scalars are not allowed at a 95.5\% probability. Since the CP-even scalar is forced to have a small mass, electroweak constraints restrict the mass spitting between the two other scalars to be small. This is clearly seen in the $M_A - M_{H^{\pm}}$ plane of Fig.~\ref{fig:MassMassH}.

Global fit results for the $\tilde{\alpha} - \varsigma_f$ planes in the positive branch are similar to the ones in the {\it light} scenario (see Fig.~\ref{fig:alpha_couplings_GF}) shifting the mixing angle $\tilde{\alpha} \to \tilde{\alpha} - \frac{\pi}{2}$.

\section{Summary}
\label{sec:summary}

Several fits to the currently available data have been performed within the A2HDM, using the \HEPfit tool that is based on Bayesian statistics. To reduce the number of fitted parameters, we have considered a CP-conserving scalar potential and real alignment couplings. We have included in the fit EWPOs, Higgs signal strengths, collider searches for additional scalars and flavour constraints, together with the theoretical requirements of perturbativity and vacuum stability. After analysing the separate implications of each type of constraints, we have combined all of them into a global fit to the model parameters.

Our main fits, discussed in Section~\ref{sec:FitResults_Light}, assume that the observed Higgs at 125~GeV is the lightest CP-even scalar of the model. The theoretical requirements strongly constrain the mass differences among the three additional scalars, $H$, $A$ and $H^\pm$, to be below 600~GeV. This upper bound becomes tighter as the masses increase, as shown in Fig.~\ref{fig:Mass_Massdiff_Lambda}. The EWPOs further restrict the masses and mass differences, favouring small mass splittings and large masses. However, the precise bounds from EWPOs turn out to be very sensitive to the adopted priors. Taking squared mass priors with masses varied until 1500~GeV, one finds a lower bound around 750 GeV for the three masses, which gets reduced to 500~GeV if a lower mass range up to 1000~GeV is adopted as prior. Taking instead linear priors in the same mass range, masses as low as 10~GeV become allowed.

The agreement of the measured Higgs signal strengths with the SM expectations translates into a very strong constraint on the scalar mixing angle, $\tilde{\alpha}\le 0.023$~rad (95.5\% probability), and tight bounds on the alignment parameters, given in Fig.~\ref{fig:alpha_couplings_S}, which are further reinforced by the flavour constraints shown in Fig.~\ref{fig:Coup_flavour_2}. Combining all constraints into a global fit, one gets finally the allowed regions shown in Figs.~\ref{fig:alpha_couplings_GF} and \ref{fig:Coup_all}.
The up-quark alignment parameter must satisfy $|\varsigma_u|< 1.5$  (95.5\% probability), while $|\varsigma_{d,\ell}|$ can take larger values provided the products $|\varsigma_u\,\varsigma_{d,\ell}|$ and $|\varsigma_{d}\,\varsigma_\ell|$ remain small. These figures assume that the down-quark and lepton Yukawa couplings do not deviate much from their SM values. However, since the current data on Higgs signal strengths cannot determine the signs of these two couplings, there is in addition a fine-tuned solution with ``wrong-sign'' Yukawas, shown in Fig.~\ref{fig:wrong_sign}.

The global fit to all data does not solve the prior dependence of the fitted mass spectrum. As shown in Figs.~\ref{fig:Mass_Massdiff_Lambda} and \ref{fig:Mass_Mass}, squared mass priors put stronger lower bounds on the scalar masses that depend on the assumed prior range, while linear priors still allow for quite low values of the masses. Clearly, the current negative results from collider searches are not yet stringent enough to discard the presence of new scalar states with masses near the electroweak scale.

We have also attempted a first study of the opposite scenario, where the 125~GeV Higgs is assumed to be the heaviest CP-even scalar. Using the same data set, we have found the constraints shown in Figs.~\ref{fig:MassMassH} and \ref{fig:St_heavy} for the masses and alignment parameters, respectively. A very strong correlation between $M_A$ and $M_{H^\pm}$ is observed at high masses, as expected, because the mass splittings cannot become large. However, no useful lower bounds on the scalar masses can be extracted because they are again too sensitive to the adopted priors. A more detailed analysis of this scenario, including LEP searches and low-energy data, could provide additional constraints that we plan to study in future works.

Our results are currently the most general global fit to the A2HDM. While previous phenomenological analyses of two-Higgs doublet models focused on particular cases based on discrete $\cZ_2$ symmetries or used only small subsets of observables, we have worked with a more generic theoretical framework with the only assumptions of flavour alignment and real scalar and alignment parameters. A more general analysis, including the new sources of CP violation provided by the A2HDM, will be attempted in future publications.

Concerning the current flavour anomalies, it is worth to compare the fitted constraints on the alignment parameters in Fig.~\ref{fig:Coup_all} with the parameter region able to accommodate $(g-2)_\mu$, shown in Fig.~\ref{fig:Coup_flavour_2}. The global fit does not exclude the large values of $|\varsigma_\ell|$ which would be needed. However, in order to fit $(g-2)_\mu$ one also needs a quite light pseudoscalar with $M_A\lesssim 50$~GeV \cite{Ilisie:2015tra,Cherchiglia:2017uwv,Wang:2018hnw}. Specific searches for light scalar and pseudoscalar particles could be very relevant to investigate this possibility. 

An explanation of the $b\to c\tau\nu$ and $b\to s\mu^+\mu^-$ anomalies within the context of two-Higgs doublet models would require non-universal lepton couplings. The generalised A2HDM \cite{Penuelas:2017ikk}, with family-dependent alignment parameters, provides a viable theoretical framework to address this type of phenomena. A scalar interpretation of the $b\to c\tau\nu$ data \cite{Celis:2012dk,Celis:2016azn} seems still possible \cite{Murgui:2019czp,Mandal:2020htr}, but it would imply higher values of $\mathrm{Br}(B_c\to\tau\nu)$ than usually assumed. A detailed analysis of the $b\to s\mu^+\mu^-$ anomalies within the generalised A2HDM would provide very useful complementary information.

\section*{Acknowledgements}

This work has been supported in part by the Spanish Government and ERDF funds from the EU Commission [grant FPA2017-84445-P], by the Generalitat Valenciana [grant Prometeo/2017/053] and by the COST Action CA16201 PARTICLEFACE. 
The research of A.P.M.\ was supported by the Cluster of Excellence {\em Precision Physics, Fundamental Interactions and Structure of Matter\/} (PRISMA${}^+$ -- EXC~2118/1) within the German Excellence Strategy (project ID 39083149). We thank Ayan Paul for his help and technical support with \HEPfit. We also thank the INFN Roma Tre Cluster, where most of the fits were performed.

\appendix
\section{Data compilation}
\label{sec:data}

The following tables detail the collider data sources employed in our global fit (data obtained at 2, 8 and 13 TeV are marked in green, purple and  yellow, respectively).
Table~\ref{tab:strengths} compiles the LHC and Tevatron data sources on Higgs signal strengths.
The information on heavy scalar searches at the LHC is collected in Tables~\ref{tab:directsearches4}, \ref{tab:directsearches1}, \ref{tab:directsearches2} and  \ref{tab:directsearches3}. 
These searches are applied either to the charged Higgs boson $H^{\pm}$ or to the neutral scalars $\varphi_i^0 = H, A$.
Direct searches related to the charged Higgs boson are displayed in Table~\ref{tab:directsearches4}.
Table~\ref{tab:directsearches1} contains information about $\varphi_i^0 = H, A$ decaying into fermions, $\gamma \gamma$ and $Z \gamma$. In Table~\ref{tab:directsearches2} the final channel is either $WW$, $ZZ$ or $VV = ZZ, WW$. Finally, information about a neutral scalar decaying into the SM Higgs boson is summarized in Table~\ref{tab:directsearches3}.   Parenthesis indicate an specific final state and square brackets that limits are quoted on the primary final state, measured through the second final state.

\begin{table}[htb]
\centering
{\renewcommand{\arraystretch}{1.}
\resizebox{\textwidth}{!}{%
\begin{tabular}{|c|c|c|c|c|c|c|c|}
\hline
\textbf{Channel} & $b \bar{b}$ & $\gamma \gamma$ & $\mu^+ \mu^-$ & $\tau^+ \tau^-$ & $WW$ & $Z \gamma $ & $ZZ$   \\
\hline
\hline
\cellcolor{color8TeV}  $\text{ggF}_8$&  & \cite{ Aad:2014eha, Khachatryan:2014ira} & \cite{ Khachatryan:2016vau} & \cite{Aad:2015vsa,  Chatrchyan:2014nva}  &  \cite{ ATLAS:2014aga, Aad:2015ona, Chatrchyan:2013iaa} & \cite{Aad:2015gba,Chatrchyan:2013vaa}  & \cite{Aad:2014eva, Khachatryan:2014jba} \\
\hline
\cellcolor{color13TeV} $\text{ggF}_{13}$&   &  \cite{Sirunyan:2018ouh,  ATLAS-CONF-2018-028} & \cite{ATLAS:2018kbw, Sirunyan:2018hbu} & \cite{Sirunyan:2017khh, Aaboud:2018pen}  &
\cite{ATLAS:2016gld, Aaboud:2018jqu, Sirunyan:2018egh}  
& \cite{Aaboud:2017uhw,Sirunyan:2018tbk}  & \cite{ATLAS:2020wny, Sirunyan:2017exp, CMS:2018mmw} \\
\hline
\hline
\cellcolor{color8TeV} $\text{VBF}_{8}$&   & \cite{ Aad:2014eha, Khachatryan:2014ira} & \cite{ Khachatryan:2016vau} & \cite{ Aad:2015vsa,  Chatrchyan:2014nva}  & \cite{ ATLAS:2014aga, Aad:2015ona, Chatrchyan:2013iaa}  & \cite{Aad:2015gba}  & \cite{Aad:2014eva, Khachatryan:2014jba} \\
\hline
\cellcolor{color13TeV} $\text{VBF}_{13}$& \cite{Aaboud:2018gay,CMS:2016mmc} &  \cite{Sirunyan:2018ouh,ATLAS-CONF-2018-028} & \cite{ATLAS:2018kbw, Sirunyan:2018hbu} & \cite{ Sirunyan:2017khh, Aaboud:2018pen} &
\cite{ATLAS:2016gld, Aaboud:2018jqu, Sirunyan:2018egh}  
& -  & \cite{ATLAS:2020wny, Sirunyan:2017exp, CMS:2018mmw} \\
\hline
\hline
\cellcolor{color8TeV} $\text{VH}_{8}$ & \cite{Aad:2014xzb, Chatrchyan:2013zna}
& \cite{ Aad:2014eha, Khachatryan:2014ira} &
\cite{ Khachatryan:2016vau} & \cite{Aad:2015vsa} & \cite{ ATLAS:2014aga,  Aad:2015ona, Chatrchyan:2013iaa} & \cite{Aad:2015gba}  & \cite{Aad:2014eva, Khachatryan:2014jba} \\
\hline
\cellcolor{color13TeV} $\text{VH}_{13}$ & \cite{Aaboud:2018zhk, Sirunyan:2017elk} & \cite{Sirunyan:2018ouh,ATLAS-CONF-2018-028} &
\cite{ATLAS:2018kbw, Sirunyan:2018hbu} & \cite{Sirunyan:2017khh, CMS:2018nqp} &
\cite{ATLAS:2016gld, Aaboud:2018jqu, Sirunyan:2018egh}  
&    & \cite{ATLAS:2020wny, Sirunyan:2017exp, CMS:2018mmw} \\
\hline
\hline
\cellcolor{color8TeV} $\text{ttH}_{8}$ & \cite{Khachatryan:2014qaa,  Aad:2015gra} & \cite{ Aad:2014eha, Khachatryan:2014ira} &\cite{Khachatryan:2016vau} &   &   & \cite{Aad:2015gba}   & \cite{Aad:2014eva, Khachatryan:2014jba} \\
\hline
\cellcolor{color13TeV} $\text{ttH}_{13}$ &  \cite{Aaboud:2017rss,CMS:2018alh,Sirunyan:2018ygk} &  \cite{Sirunyan:2018ouh,ATLAS-CONF-2018-028} &\cite{ATLAS:2018kbw, Sirunyan:2018hbu} &  & \cite{Sirunyan:2018egh}   &     & \cite{ATLAS:2020wny, Sirunyan:2017exp, CMS:2018mmw} \\
\hline
\hline
\cellcolor{color2TeV} $\text{VH}_{2}$ &  \cite{Abazov:2013gmz,Aaltonen:2013ipa} &   &
  &   &   &   &   \\
\hline
\cellcolor{color2TeV} $\text{ttH}_{2}$ &\cite{ Abazov:2013gmz,Aaltonen:2013ipa} &   &   &   &   &     &   \\
\hline
\end{tabular}
}
}
\caption{Higgs signal strengths input used in the fit, for different production and decay channels, at energies of $\sqrt{s} = 7,8 $ TeV (ATLAS and CMS, Run I), $\sqrt{s} = 13 $ TeV (ATLAS and CMS, Run II) and $\sqrt{s} = 2 $ TeV (D0 and CDF collaborations).}
\label{tab:strengths}
\end{table}

\begin{table}[htb]
\begin{center}
\begin{tabular}{| l | l | l c | c | c |}
\hline
\textbf{Label} &\textbf{Channel} & \multicolumn{2}{| l |}{\textbf{Experiment}} & \textbf{Mass range} & ${\cal L}$ \\
&&&& \textbf{[TeV]} & \textbf{[fb$^{-1}$]} \\[1pt]
\hline
\cellcolor{color8TeV} $A_{8}^{\tau\nu}$ & $pp\to H^\pm \to \tau^\pm \nu $ & ATLAS &\cite{Aad:2014kga} & [0.18;1] & 19.5\\
\hline
\cellcolor{color8TeV} $C_{8}^{\tau\nu}$  & $pp\to H^+ \to \tau^+ \nu $ & CMS &\cite{Khachatryan:2015qxa}& [0.18;0.6] & 19.7\\
\hline
\cellcolor{color13TeV} $A_{13}^{\tau\nu}$  & \multirow{2}{*}{$pp\to H^{\pm} \to \tau^\pm \nu $ }& ATLAS & \cite{Aaboud:2018gjj} & [0.09;2] & 36.1\\
\cellcolor{color13TeV} $C_{13}^{\tau\nu}$ & & CMS & \cite{CMS-PAS-HIG-16-031} & [0.18;3] & 12.9\\
\hline
\hline
\cellcolor{color8TeV} $A_{8}^{tb}$ & $pp\to H^\pm \to t b $ & ATLAS & \cite{Aad:2015typ} & [0.2;0.6] & 20.3\\
\hline
\cellcolor{color8TeV} $C_{8}^{tb}$ & $pp\to H^+ \to t \bar{b} $ & CMS & \cite{Khachatryan:2015qxa} & [0.18;0.6] & 19.7\\
\hline
\cellcolor{color13TeV} $A_{13}^{tb}$ & $pp\to H^\pm \to tb $ & ATLAS & \cite{Aaboud:2018cwk} & [0.2;2] & 36.1\\
\hline
\end{tabular}
\caption{Direct searches for charged  scalars.}
\label{tab:directsearches4}
\end{center}
\end{table}

\begin{table}[htb]
\centering
{\renewcommand{\arraystretch}{1.}
\resizebox{\textwidth}{!}{%
\begin{tabular}{|l|l|lc|c|c|}
\hline
\textbf{Label} &\textbf{Channel} & \multicolumn{2}{| l |}{\textbf{Experiment}} & \textbf{Mass range} & ${\cal L}$ \\
&&&& \textbf{[TeV]} & \textbf{[fb$^{-1}$]} \\[1pt]
\hline
\hline
\cellcolor{color13TeV} $A_{13t}^{tt}$  & $tt \to \varphi_i^0 \to tt$ & ATLAS & \cite{Aaboud:2018xpj} & [0.4;1] & 36.1 \\
\hline
\cellcolor{color13TeV} $A_{13b}^{tt}$  & $bb \to \varphi_i^0 \to tt$ & ATLAS & \cite{ATLAS-CONF-2016-104} & [0.4;1] & 13.2\\
\hline
\hline
\cellcolor{color8TeV} $C_{8b}^{bb}$ &$bb \to \varphi_i^0 \to bb$ & CMS & \cite{Khachatryan:2015tra} & [0.1;0.9] & 19.7\\
\hline
\cellcolor{color8TeV} $C_{8}^{bb}$  &$gg \to \varphi_i^0\to bb$ & CMS & \cite{Sirunyan:2018pas} & [0.33;1.2] & 19.7\\
\hline
\cellcolor{color13TeV} $C_{13}^{bb}$ & $pp \to \varphi_i^0\to bb$ & CMS & \cite{CMS-PAS-HIG-16-025} & [0.55;1.2] & 2.69\\
\hline
\cellcolor{color13TeV} $C_{13b}^{bb}$ & $bb \to \varphi_i^0\to bb$ & CMS & \cite{Sirunyan:2018taj} & [0.3;1.3] & 35.7\\
\hline
\hline
\cellcolor{color8TeV} $A_{8}^{\tau\tau}$ &\multirow{2}{*}{$gg\to \varphi_i^0 \to \tau\tau$} & ATLAS &\cite{Aad:2014vgg} & [0.09;1] & 20 \\
\cellcolor{color8TeV} $C_{8}^{\tau\tau}$  & & CMS &\cite{CMS-PAS-HIG-14-029} &  [0.09;1]  &19.7 \\
\hline
\cellcolor{color8TeV} $A_{8b}^{\tau\tau}$  &\multirow{2}{*}{$bb\to \varphi_i^0 \to \tau\tau$} & ATLAS &\cite{Aad:2014vgg} & [0.09;1] & 20 \\
\cellcolor{color8TeV} $C_{8b}^{\tau\tau}$ & & CMS & \cite{CMS-PAS-HIG-14-029}& [0.09;1] & 19.7 \\
\hline
\cellcolor{color13TeV} $A_{13}^{\tau\tau}$ &\multirow{2}{*}{$gg \to \varphi_i^0\to \tau \tau$} & ATLAS & \cite{Aaboud:2017sjh} & [0.2;2.25] & 36.1\\[-1pt]
\cellcolor{color13TeV} $C_{13}^{\tau\tau}$ &  & CMS & \cite{Sirunyan:2018zut} & [0.09;3.2] & 35.9\\
\hline
\cellcolor{color13TeV} $A_{13b}^{\tau\tau}$ &\multirow{2}{*}{$bb \to \varphi_i^0\to \tau \tau$} & ATLAS & \cite{Aaboud:2017sjh} & [0.2;2.25] & 36.1\\[-1pt]
\cellcolor{color13TeV} $C_{13b}^{\tau\tau}$ & & CMS & \cite{Sirunyan:2018zut} & [0.09;3.2] & 35.9\\
\hline
\hline
\cellcolor{color8TeV} $A_{8}^{\gamma\gamma}$&$gg\to \varphi_i^0 \to \gamma\gamma$ & ATLAS &\cite{Aad:2014ioa} & [0.065;0.6] & 20.3 \\
\hline
\cellcolor{color13TeV} $A_{13}^{\gamma\gamma}$ &$pp \to \varphi_i^0\to \gamma \gamma$ & ATLAS & \cite{Aaboud:2017yyg} & [0.2;2.7] & 36.7\\
\hline
\cellcolor{color13TeV} $C_{13}^{\gamma\gamma}$ & $gg \to \varphi_i^0\to \gamma \gamma$ & CMS & \cite{Khachatryan:2016yec} & [0.5;4] & 35.9\\ 
\hline
\hline
\cellcolor{color8TeV} $A_{8}^{Z\gamma}$  &\multirow{2}{*}{$pp\to \varphi_i^0 \to Z\gamma \to (\ell \ell) \gamma$} & ATLAS & \cite{Aad:2014fha} & [0.2;1.6] & 20.3 \\
\cellcolor{color8TeV} $C_{8}^{Z\gamma}$ & & CMS & \cite{CMS-PAS-HIG-16-014} & [0.2;1.2] & 19.7 \\
\hline
\cellcolor{color13TeV} $A_{13}^{\ell\ell\gamma}$  & $gg \to \varphi_i^0\to Z \gamma [\to (\ell \ell) \gamma ]$ & ATLAS & \cite{Aaboud:2017uhw} & [0.25;2.4] & 36.1\\
\hline
\cellcolor{color13TeV} $A_{13}^{qq\gamma}$  & $gg \to \varphi_i^0\to Z \gamma [\to (qq) \gamma ]$ & ATLAS & \cite{Aaboud:2018fgi} & [1;6.8] & 36.1\\
\hline
\cellcolor{color13TeV} $C_{8+13}^{Z\gamma}$  & $gg \to \varphi_i^0\to Z \gamma$ & CMS & \cite{Sirunyan:2017hsb} & [0.35;4] & 35.9\\
\hline
\end{tabular}
}
}
\caption{Direct searches for neutral heavy scalars, $\varphi_i^0 = H, A$, with quarks, leptons ($\ell= e,\mu$), photons and $Z\gamma$ final states.}
\label{tab:directsearches1}
\end{table}

\begin{table}[htb]
\centering
{\renewcommand{\arraystretch}{1.}
\resizebox{\textwidth}{!}{%
\begin{tabular}{|l|l|lc|c|c|}
\hline
\textbf{Label} &\textbf{Channel} & \multicolumn{2}{| l |}{\textbf{Experiment}} & \textbf{Mass range} & ${\cal L}$ \\
&&&& \textbf{[TeV]} & \textbf{[fb$^{-1}$]} \\[1pt]
\hline
\cellcolor{color8TeV} $A_{8}^{ZZ}$  &$gg\to \varphi_i^0\to ZZ$ & ATLAS & \cite{Aad:2015kna}& [0.14;1] & 20.3 \\
\hline
\cellcolor{color8TeV} $A_{8V}^{ZZ}$  &$VV \to \varphi_i^0\to ZZ$ & ATLAS & \cite{Aad:2015kna}& [0.14;1] & 20.3 \\
\hline
\cellcolor{color13TeV} $A_{13}^{2\ell2L}$  & $gg\to \varphi_i^0 \to ZZ [\to (\ell \ell) (\ell \ell, \nu \nu)]$ & ATLAS & \cite{Aaboud:2017rel} & [0.2;1.2] & 36.1\\
\hline
\cellcolor{color13TeV} $A_{13V}^{2\ell2L}$  & $VV\to \varphi_i^0\to ZZ [\to (\ell \ell) (\ell \ell, \nu \nu)]$ & ATLAS & \cite{Aaboud:2017rel} & [0.2;1.2] & 36.1\\
\hline
\cellcolor{color13TeV} $A_{13}^{2L2q}$ & $gg\to \varphi_i^0\to ZZ [\to (\ell \ell, \nu \nu) (qq)]$ & ATLAS & \cite{Aaboud:2017itg} & [0.3;3] & 36.1\\
\hline
\cellcolor{color13TeV} $A_{13V}^{2L2q}$ & $VV\to \varphi_i^0\to ZZ [\to (\ell \ell, \nu \nu) (qq)]$ & ATLAS & \cite{Aaboud:2017itg} & [0.3;3] & 36.1\\
\hline
\cellcolor{color13TeV} $C_{13}^{2\ell2X}$ & $pp\to \varphi_i^0\to ZZ [\to (\ell \ell) (qq,\nu\nu,\ell\ell)]$ & CMS & \cite{Sirunyan:2018qlb} & [0.13;3] & 35.9\\
\hline
\cellcolor{color13TeV} $C_{13}^{2q2\nu}$ & $pp\to \varphi_i^0\to ZZ [\to (qq)(\nu\nu)]$ & CMS & \cite{Sirunyan:2018ivv} & [1;4] & 35.9\\
\hline
\hline
\cellcolor{color8TeV} $A_{8}^{WW}$  &$gg\to \varphi_i^0\to WW$ & ATLAS &\cite{Aad:2015agg}& [0.3;1.5] & 20.3 \\
\hline
\cellcolor{color8TeV} $A_{8V}^{WW}$ &$VV \to \varphi_i^0\to WW$ & ATLAS & \cite{Aad:2015agg}& [0.3;1.5] & 20.3 \\
\hline
\cellcolor{color13TeV} $A_{13}^{2(\ell\nu)}$  & $gg\to \varphi_i^0\to WW [\to (e \nu) (\mu \nu)]$ & ATLAS & \cite{Aaboud:2017gsl} & [0.2;4] & 36.1\\
\hline
\cellcolor{color13TeV} $A_{13V}^{2(\ell\nu)}$  & $VV\to \varphi_i^0\to WW [\to (e \nu) (\mu \nu)]$ & ATLAS & \cite{Aaboud:2017gsl} & [0.2;3] & 36.1\\
\hline
\cellcolor{color13TeV} $C_{13}^{2(\ell\nu)}$ & $(gg\!+\!VV)\to \varphi_i^0\to WW \to (\ell \nu) (\ell \nu)$ & CMS & \cite{CMS-PAS-HIG-16-023} & [0.2;1] & 2.3\\
\hline
\cellcolor{color13TeV} $A_{13}^{\ell\nu2q}$ & $gg\to \varphi_i^0\to WW[\to (\ell \nu) (qq)]$ & ATLAS & \cite{Aaboud:2017fgj} & [0.3;3] & 36.1\\
\hline
\cellcolor{color13TeV} $A_{13V}^{\ell\nu2q}$ & $VV\to \varphi_i^0\to WW[\to (\ell \nu) (qq)]$ & ATLAS & \cite{Aaboud:2017fgj} & [0.3;3] & 36.1\\
\hline
\cellcolor{color13TeV} $C_{13}^{\ell\nu2q}$ & $pp\to \varphi_i^0\to WW[\to (\ell \nu) (qq)]$ & CMS & \cite{Sirunyan:2018iff} & [1;4.4] & 35.9\\
\hline
\hline
\cellcolor{color8TeV} $C_{8}^{VV}$  & $pp \to \varphi_i^0\to VV$ & CMS & \cite{Khachatryan:2015cwa} & [0.145;1] & 24.8 \\
\cellcolor{color13TeV} $A_{13}^{4q}$  & $pp\to \varphi_i^0\to VV [\to (qq) (qq)]$ & ATLAS & \cite{Aaboud:2017eta} & [1.2;3] & 36.7\\
\hline
\end{tabular}
}
}
\caption{Direct searches for neutral heavy scalars, $\varphi_i^0 = H, A$, with vector-boson final states. $V = W,Z$, $\ell = e, \mu$.}
\label{tab:directsearches2}
\end{table}

\begin{table}[htb]
\centering
{\renewcommand{\arraystretch}{0.9}
\resizebox{\textwidth}{!}{%
\begin{tabular}{| l | l | l c | c | c |}
\hline
\textbf{Label} &\textbf{Channel} & \multicolumn{2}{| l |}{\textbf{Experiment}} & \textbf{Mass range} & ${\cal L}$ \\
&&&& \textbf{[TeV]} & \textbf{[fb$^{-1}$]} \\[1pt]
\hline
\hline
\cellcolor{color8TeV} $A_{8}^{hh}$ &$gg\to \varphi_i^0 \to hh$ & ATLAS &\cite{Aad:2015xja} & [0.26;1] & 20.3\\
\hline
\cellcolor{color8TeV} $C_{8}^{4b}$  &$pp\to  \varphi_i^0 \to hh \to (bb) (bb)$ & CMS &\cite{Khachatryan:2015yea} & [0.27;1.1] & 17.9\\
\hline
\cellcolor{color8TeV} $C_{8}^{2\gamma2b}$  &$pp\to  \varphi_i^0 \to hh \to (bb) (\gamma \gamma)$ & CMS & \cite{Khachatryan:2016sey} & [0.260;1.1] & 19.7\\
\hline
\cellcolor{color8TeV} $C_{8g}^{2b2\tau}$ &$gg\to  \varphi_i^0 \to hh \to (bb) (\tau\tau)$ & CMS & \cite{Khachatryan:2015tha} & [0.26;0.35] & 19.7\\
\hline
\cellcolor{color8TeV} $C_{8}^{2b2\tau}$  &$pp\to  \varphi_i^0 \to hh [\to (bb) (\tau\tau)]$ & CMS & \cite{Sirunyan:2017tqo} & [0.35;1] & 18.3\\
\hline
\cellcolor{color13TeV} $A_{13}^{4b}$& \multirow{3}{*}{$pp \to  \varphi_i^0 \to hh \to (bb) (bb)$} & ATLAS & \cite{Aaboud:2018knk} & [0.26;3]  & 36.1\\[-1pt]
\cellcolor{color13TeV} $C_{13,1}^{4b}$  & & CMS & \cite{Sirunyan:2018zkk} & [0.26;1.2] & 35.9\\
\cellcolor{color13TeV} $C_{13,2}^{4b}$  & & CMS & \cite{Sirunyan:2018qca} & [1.2;3] & 35.9\\
\hline
\cellcolor{color13TeV} $A_{13}^{2\gamma2b}$  & $pp \to  \varphi_i^0 \to hh [\to (bb) (\gamma \gamma)]$ & ATLAS & \cite{Aaboud:2018ftw} & [0.26;1] & 36.1\\
\cellcolor{color13TeV} $C_{13}^{2\gamma2b}$  & $pp \to  \varphi_i^0 \to hh \to (bb) (\gamma \gamma)$ & CMS & \cite{Sirunyan:2018iwt} & [0.25;0.9] & 35.9\\
\hline
\cellcolor{color13TeV} $A_{13}^{2b2\tau}$ & \multirow{2}{*}{$pp \to \varphi_i^0 \to hh \to (bb) (\tau \tau)$} & ATLAS & \cite{Aaboud:2018sfw} & [0.26;1] & 36.1\\
\cellcolor{color13TeV} $C_{13,1}^{2b2\tau}$ & & CMS & \cite{Sirunyan:2017djm} & [0.25;0.9] & 35.9\\
\cellcolor{color13TeV} $C_{13,2}^{2b2\tau}$ & $pp \to  \varphi_i^0 \to hh [\to (bb) (\tau \tau)]$ & CMS & \cite{Sirunyan:2018fuh} & [0.9;4] & 35.9\\
\hline
\cellcolor{color13TeV} $C_{13}^{2b2V}$ & $pp \to  \varphi_i^0 \to hh \to (bb) (VV\to \ell \nu \ell \nu)$ & CMS & \cite{Sirunyan:2017guj} & [0.26;0.9] & 35.9\\
\hline
\cellcolor{color13TeV} $A_{13}^{2b2W}$ & $pp \to  \varphi_i^0 \to hh [\to (bb) (WW)]$ & ATLAS & \cite{Aaboud:2018zhh} & [0.5;3] & 36.1\\
\hline
\cellcolor{color13TeV} $A_{13}^{2\gamma2W}$  & $gg \to  \varphi_i^0 \to hh \to (\gamma \gamma) (WW)$ & ATLAS & \cite{Aaboud:2018ewm} & [0.26;0.5] & 36.1\\ 
\hline
\hline
\cellcolor{color8TeV} $A_{8}^{bbZ}$ &$gg\to  \varphi_i^0 \to hZ \to (bb) Z$ & ATLAS & \cite{Aad:2015wra} & [0.22;1] & 20.3\\
\hline
\cellcolor{color8TeV} $C_{8}^{2b2\ell}$ &$gg\to  \varphi_i^0 \to hZ \to (bb) (\ell \ell)$ & CMS &\cite{Khachatryan:2015lba} & [0.225;0.6] &19.7\\
\hline
\cellcolor{color8TeV} $A_{8}^{\tau\tau Z}$ &$gg\to  \varphi_i^0 \to hZ \to (\tau\tau) Z$ & ATLAS & \cite{Aad:2015wra} & [0.22;1] & 20.3\\
\hline
\cellcolor{color8TeV} $C_{8}^{2\tau2\ell}$  &$gg\to  \varphi_i^0 \to hZ \to (\tau\tau) (\ell \ell)$ & CMS & \cite{Khachatryan:2015tha} & [0.22;0.35] & 19.7\\
\hline
\cellcolor{color13TeV} $A_{13}^{bbZ}$  & \multirow{3}{*}{$gg\to  \varphi_i^0 \to hZ \to (bb) Z$} & ATLAS & \cite{Aaboud:2017cxo} & [0.2;2] & 36.1\\
\cellcolor{color13TeV} $C_{13,1}^{bbZ}$  & & CMS & \cite{Sirunyan:2019xls} & [0.22;0.8] & 35.9\\
\cellcolor{color13TeV} $C_{13,2}^{bbZ}$  & & CMS & \cite{Sirunyan:2018qob} & [0.8;2] & 35.9\\
\hline
\cellcolor{color13TeV} $A_{13b}^{bbZ}$  & \multirow{3}{*}{$bb\to  \varphi_i^0 \to hZ \to (bb) Z$} & ATLAS & \cite{Aaboud:2017cxo} & [0.2;2] & 36.1\\
\cellcolor{color13TeV} $C_{13b,1}^{bbZ}$  & & CMS & \cite{Sirunyan:2019xls} & [0.22;0.8] & 35.9\\
\cellcolor{color13TeV} $C_{13b,2}^{bbZ}$  & & CMS & \cite{Sirunyan:2018qob} & [0.8;2] & 35.9\\
\hline
\hline
\cellcolor{color8TeV} $C_{8, 1}^{\varphi_2^0 Z}$& $pp\to \varphi_3^0  \to \varphi_2^0 Z \to (bb) (\ell\ell)$ & CMS & \cite{Khachatryan:2016are} & [0.04;1] & 19.8\\
\hline
\cellcolor{color8TeV} $C_{8, 2}^{\varphi_2^0 Z}$& $pp\to \varphi_3^0 \to \varphi_2^0 Z \to (\tau \tau) (\ell\ell)$ & CMS & \cite{Khachatryan:2016are} & [0.05;1] & 19.8\\
\hline
\cellcolor{color13TeV} $A_{13}^{\varphi^0 Z}$& $gg\to \varphi_{3}^0 \to \varphi_2^0 Z \to (bb) Z$ & ATLAS & \cite{Aaboud:2018eoy} & [0.13;0.8] & 36.1\\
\hline
\cellcolor{color13TeV} $A_{13b}^{\varphi^0 Z}$& $bb\to \varphi_{3}^0 \to \varphi_2^0 Z \to (bb) Z$ & ATLAS & \cite{Aaboud:2018eoy} & [0.13;0.8] & 36.1\\
\hline
\end{tabular}
}
}
\caption{Direct searches for neutral heavy scalars, $\varphi_i^0 = H, A$, with final states including the SM Higgs boson or other neutral scalars. $\varphi_3$ denotes the heaviest scalar,  $V = W,Z$, $\ell = e, \mu$.}
\label{tab:directsearches3}
\end{table}

\mbox{}\vskip 1cm


\clearpage

\bibliographystyle{JHEP}
\bibliography{ATHDM_fit.bib}

\end{document}